\documentclass[a4paper,fleqn,usenatbib]{mnras}

\usepackage{newtxtext,newtxmath}

\usepackage[T1]{fontenc}
\usepackage{ae,aecompl}

\usepackage{multicol,graphicx,longtable,float,caption,adjustbox,rotating,datetime,geometry,wrapfig,rotating,threeparttablex,booktabs}
\usepackage{amsmath}

\newcommand{\myfnmark}[1]{\mbox{\textsuperscript{\normalfont #1}}}

\title[HMXBs in M33]{Formation Timescales for High-Mass X-ray Binaries in M33}

\author[K. Garofali et al.]{
Kristen Garofali,$^{1}$\thanks{E-mail: garofali@uw.edu (KG)}
Benjamin F. Williams,$^{1}$
Tristan Hillis,$^{1}$
Karoline M. Gilbert,$^{2}$ 
\newauthor{Andrew E. Dolphin,$^{3}$}
Michael Eracleous,$^{4}$
Breanna Binder$^{5}$
\\
$^{1}$ Astronomy Department, University of Washington, Box 351580, Seattle, WA, 98195; garofali@uw.edu; ben@astro.washington.edu; \\ tristan3214@live.com\\
$^{2}$ Space Telescope Science Institute, 3700 San Martin Dr., Baltimore, MD 21218, USA; kgilbert@stsci.edu \\
$^{3}$ Raytheon Company, Tucson, AZ 85734, USA; adolphin@raytheon.com\\
$^{4}$ Department of Astronomy \& Astrophysics and Institute for Gravitation and the Cosmos, The Pennsylvania State University, \\ 525 Davey Lab, University Park, PA, 16802; mxe17@psu.edu\\
$^{5}$ Department of Physics \& Astronomy, California State Polytechnic University, 3801 West Temple Ave., Pomona, CA 91768; babinder@cpp.edu}

\date{Accepted 2018 June 14. Received 2018 June 14; in original form 2018 January 6}

\pubyear{2018}

\begin{document}
\label{firstpage}
\pagerange{\pageref{firstpage}--\pageref{lastpage}}
\maketitle

\begin{abstract}

We have identified 55 candidate high-mass X-ray binaries (HMXBs) in M33 using available archival {\it HST} and {\it Chandra} imaging to find blue stars associated with X-ray positions. We use the {\it HST} photometric data to model the color-magnitude diagrams in the vicinity of each candidate HMXB to measure a resolved recent star formation history (SFH), and thus a formation timescale, or age for the source. Taken together, the SFHs for all candidate HMXBs in M33 yield an age distribution that suggests preferred formation timescales for HMXBs in M33 of $<$ 5 Myr and $\sim$ 40 Myr after the initial star formation episode. The population at 40 Myr is seen in other Local Group galaxies, and can be attributed to a peak in formation efficiency of HMXBs with neutron stars as compact objects and B star secondary companions. This timescale is preferred as neutron stars should form in abundance from $\sim$ 8 M$_{\odot}$ core-collapse progenitors on these timescales, and B stars are shown observationally to be most actively losing mass around this time. The young population at $<$ 5 Myr has not be observed in other Local Group HMXB population studies, but may be attributed to a population of very massive progenitors forming black holes very early on. We discuss these results in the context of massive binary evolution, and the implications for compact object binaries and gravitational wave sources. 
 
\end{abstract}




\section{Introduction}

Massive stars are key drivers in the cycle of star formation, feedback, galactic chemical evolution, and compact object formation, but many aspects of the evolution of high-mass stars, in particular the end stages of their lifetimes, are still difficult to constrain and model. Furthermore, the majority of massive stars ($\sim$ 70\%) are in close enough binaries for mass transfer or mergers to occur, which can dramatically alter the evolution of the stars \citep[e.g.][]{Sana2012,deMink2013}. The evolutionary paths taken depend strongly on often-unknown initial orbital parameters, and suffer from uncertainties in prescriptions for mass-loss rates, metallicity effects, mass transfer efficiency, the common-envelope phase, and more.

The favored evolutionary pathways and uncertainties in prescriptions for mass transfer efficiency and the common envelope phase can be mitigated by observing and modeling sources at various evolutionary stages. High-mass X-ray binaries (HMXBs), systems containing a compact object (black hole or neutron star) and a high-mass ($>$ $\sim$ 8 M$_{\odot}$) star with close enough orbital separation such that accretion onto the compact object results in an X-ray bright source, offer a unique window into massive binary evolution as they can act as a fossil record of past binary interactions. In addition, HMXBs may be gamma-ray burst progenitors \citep{dominik2012}, precursors to close compact object binaries and gravitational wave sources \citep{Hainich2017, Tauris2017}, and potentially play a role in heating the intergalactic medium at high redshift \citep{Justham2012,Fragos2013,Madau2017}. Therefore, constraining the observed HMXB population across a variety of environments is of utmost importance to a broad range of astrophysics. 

Studies of resolved HMXB populations in the Milky Way and Local Group galaxies have accumulated hundreds of candidate sources that have been used to constrain and confirm many key evolutionary qualities, such as neutron star (NS) kick velocities, the formation rate of X-ray binaries (XRBs) locally, the X-ray luminosity function (XLF) shape, correlations with host galaxy star formation rate (SFR), and the relation of HMXB formation to parent stellar population age \citep[e.g.][]{Liu2000,Grimm2003,SG2005,Coe2005,Liu2006,Antoniou2010,Binder2012,Mineo2012,Sturm2013,Williams2013,Walter2015,Antoniou2016,Haberl2016,Laycock2017}. 

In resolved HMXB population studies it is instructive to delineate between the various HMXB subclasses. The most numerous observationally derived subclass of HMXBs are Be/X-ray binaries (Be-XRBs), systems that typically consist of a NS in a high eccentricity orbit that accretes material as it passes through the decretion disk of its companion. The companion stars are more easily identified in this case by the strong Balmer emission emanating from their equatorial disks \citep{Porter2003}. By contrast, there is a relative paucity of observed HMXBs with supergiant companions (SG-XRBs), which are typically wind-fed systems. There is only one such system known in the SMC \citep{Coe2005}, four systems in the LMC \citep{Antoniou2016}, and an increasing number in the Milky Way thanks to observations with {\it INTEGRAL} \citep[e.g.][]{Walter2015}. There may also be a large population of  unobserved HMXBs with supergiant companions belonging to the more recently discovered class of systems known as supergiant fast X-ray transients (SFXTs). This subclass is generally harder to detect owing to the systems' extremely short outbursts (minutes to hours), and low luminosities in quiescence \citep{Negueruela2006}.

A notable exception to recent HMXB population studies in the Local Group is M33, which is an excellent target in which to study the HMXB population using a combination of {\it Chandra} and the {\it Hubble Space Telescope} ({\it HST}). M33 does not suffer from the same distance and extinction uncertainties that hinder galactic samples, and it has the deepest and most spatially resolved {\it Chandra} and {\it XMM-Newton} surveys to date of any nearby spiral \citep[ChASeM33 survey,][]{Tullmann2011,Williams2015} coupled with extensive archival {\it HST} coverage. To date, there are still only three HMXBs that 
have been robustly characterized in M33, and all are of great interest: the nucleus of M33 \citep{Long2002}, and two X-ray eclipsing binaries \citep{Pietsch2004, Pietsch2006, Pietsch2009}. Of these, M33 X-7 hosts one of the most massive stellar mass black holes (15.65~M$_{\odot}$) observed pre-LIGO \citep{Orosz2007}. 

Recent analysis of the M33 XLF reveals a slope that is consistent with the universal slope expected for a population of HMXBs \citep{Mineo2012}, suggesting that the M33 source population is dominated by HMXBs \citep{Williams2015}. The normalization of the XLF, and therefore number of HMXBs expected, is related to the SFR of the host galaxy \citep{Grimm2003,Gilfanov2004,Mineo2012}. We can therefore provide a rough estimate of the expected size of the HMXB population in M33 using the known SFR of M33 and the N$_\textrm{\scriptsize HMXB}$--SFR relation from \cite{Grimm2003}. Adopting a SFR of 0.3 M$_{\odot}$~yr$^{-1}$ for M33 \citep{WilliamsSFR} we can infer that there are likely $\sim$ 109 HMXBs in M33 assuming a limiting luminosity of $\sim$ 1 $\times$ 10$^{35}$ ergs~s$^{-1}$. This suggests that there are an order of magnitude more HMXBs in M33 yet to be discovered, characterized, and compared to models of massive binary formation and evolution.

In this paper, we present work on the identification of 55 HMXB candidates in M33 via a combination of archival {\it Chandra} and {\it HST} data. We further measure the age distribution for this HMXB population using resolved star formation histories (SFHs). In Section~\ref{catalogs} we describe the catalogs used in this analysis, the image alignment and SFH recovery technique, and the HMXB classification scheme. In Section~\ref{results} we describe the population of candidate HMXBs in M33 identified in this work, and discuss the age measurements for individual sources of interest, as well as the entire population. In Section~\ref{discuss} we discuss the effects of spurious sources on the results, how the HMXB age distribution compares to measurements of HMXB ages in other Local Group galaxies, models of massive binary evolution, and the implications for compact object binary formation. Finally, in Section~\ref{conclude} we present our conclusions.

\section{Data Acquisition \& Analysis Techniques}\label{catalogs}

In this section, we describe the catalogs used in this analysis, namely {\it Chandra} data from the {\it Chandra} ACIS Survey of M33 \citep[ChASeM33,][hereafter T11]{Tullmann2011}, for localizing X-ray point sources, archival {\it HST} data for identifying optical counterparts to X-ray sources, and the {\it Spitzer} catalog of \citet{Khan2015} for cross-correlating with our X-ray catalog to refine source characterization. In addition, we describe the technique used to align the {\it Chandra} and {\it HST} data to common frame so that we can determine candidate optical counterparts to X-ray sources within {\it Chandra} error circles. We then describe how this multi-wavelength coverage can be leveraged to find candidate HMXBs in M33. Finally, we discuss the color magnitude diagram (CMD) fitting technique used to recover resolved SFHs, and thus ages, in the vicinities of HMXBs.

\subsection{{\it Chandra} Catalog}\label{chandra}

All X-ray sources used in this analysis come from the high resolution ChASeM33 survey (T11), which had a total exposure time of 1.4 Ms and covered about 70\% of the D$_{25}$ isophote of M33 down to a limiting 0.35-8.0 keV luminosity of 2.4$\times$10$^{34}$ ergs~s$^{-1}$. This catalog contains 662 X-ray sources for which there are positions that have been aligned to 2MASS \citep{Cutri2003}, fluxes, hardness ratios (HRs), variability information, and spectral information for the subset containing enough counts. Based on the analysis of T11 $\sim$ 200 of the sources in their catalog were assigned tentative classifications.

\begin{figure}
\centering
\includegraphics[width=0.5\textwidth,trim=20 0 0 0, clip]{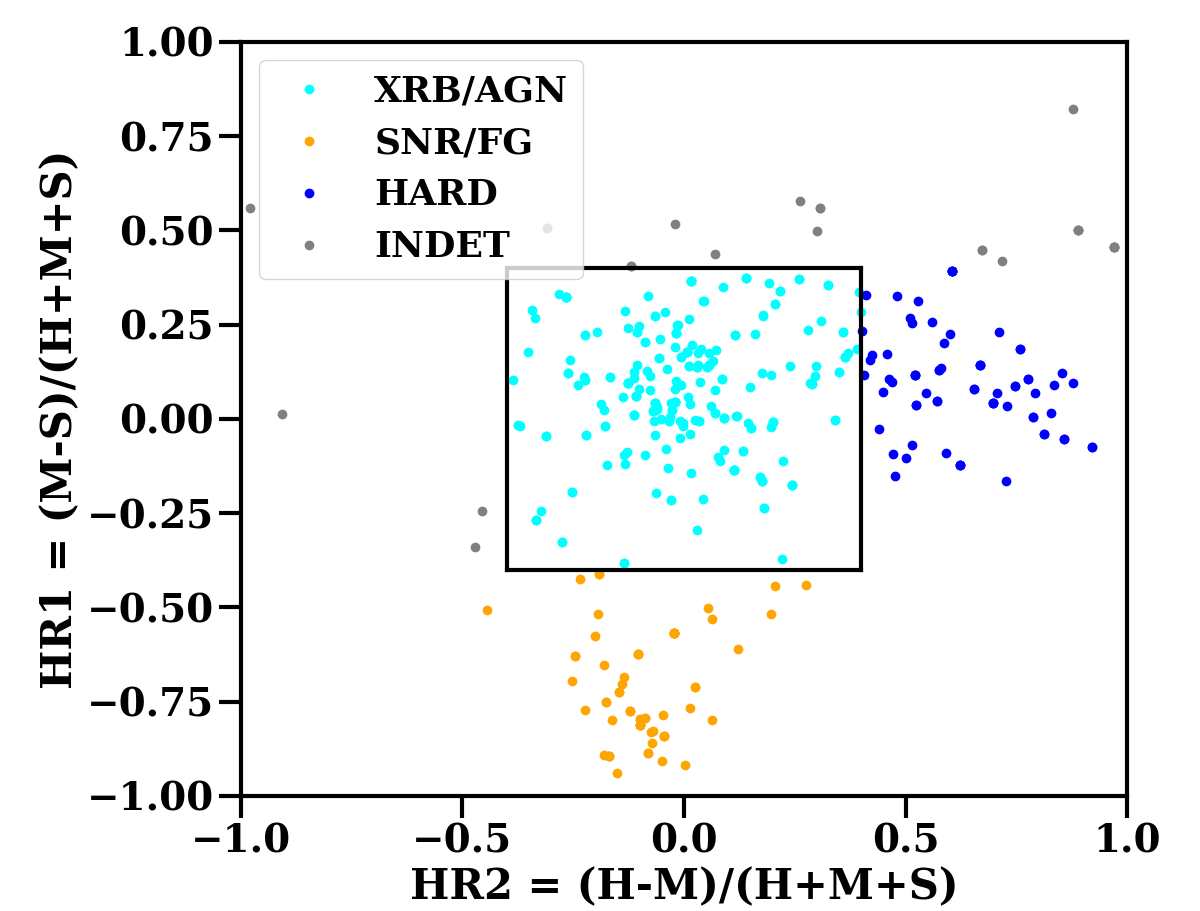}
\caption{Hardness ratios for sources from the T11 catalog in the soft (0.35-1.1 keV), medium (1.1-2.6 keV), and hard (2.6-8.0 keV) bands. Sources are color-coded and labeled by their likely type based on HRs. AGN and XRBs (both LMXBs and HMXBs) are in cyan and occupy the same locus outlined in black. Soft sources (SNRs and foreground stars) are in orange, while hard sources (e.g. those behind large column densities) are in blue. Sources of indeterminate type based on HRs are in grey.} \label{hr_box}
\end{figure}

In this analysis we primarily used the X-ray source positions from T11 for localizing candidates of interest, in addition to HRs, variability information, and past classifications to narrow down the catalog of HMXBs in M33. The HRs in particular were used for separating soft sources, foreground stars and supernova remnants (SNRs), from harder sources like background active galactic nuclei (AGN) and XRBs, both HMXBs and low-mass X-ray binaries (LMXBs). An example of HR cuts used to roughly categorize sources is shown in Figure~\ref{hr_box} using HR cuts from \citet{Binder2012}. The energy bands from T11 used are 0.35-1.1 keV (S), 1.1-2.6 keV (M), and 2.6-8.0 keV (H). AGN and XRBs occupy the same locus in the HR plot, outlined in black, while soft sources, and harder (typically absorbed) sources are found outside this region.  

We further characterized sources using the source variability flags from T11 designed to separate non-variable from both short- and long-term variable sources. Sources like SNRs should not be highly variable in time, while XRBs, AGN and foreground stars can be variable on both short and long timescales. In particular, AGN may be variable on short timescales, of order hours and possibly shorter \citep{Moran2005,Shu2017,HernandezGarcia2017}, while XRBs may be variable on extremely short (e.g. millisecond) timescales to relatively long timescales for systems that display aperiodic outbursts \citep{vdk2004}. Thus sources that display rapid variability are potential HMXB candidates, while those sources with longer term variability may be AGN or XRBs, which can further be distinguished by their candidate optical counterpart colors where {\it HST} data is available. 

In this analysis we also used the pre-determined source classifications from T11 that were made on the basis of cross-correlation with pre-existing catalogs to determine potential optical counterparts, and X-ray and optical spectra (where available).  The T11 catalog was cross-correlated with the {\it XMM-Newton} catalogs of \citet{PMH2004}
and \citet{Misanovic2006}, the SNR catalog of \citet{Long2010}, the 2MASS All-Sky point source catalog \citep{Cutri2003}, the USNO-B1.0 catalog \citep{Monet2003}, and most recently with the deep {\it XMM-Newton} catalog of M33 from \citet{Williams2015} for which there are X-ray HRs and X-ray timing analysis. In addition, the T11 catalog contains X-ray spectral fits for $\sim$ 250 sources using the {\it Chandra} data, as well as optical spectroscopy for 116 sources using Hectospec on the MMT. For sources without Hectospec spectra T11 utilized archival {\it Spitzer}, {\it GALEX} and Local Group Galaxy Survey \citep[LGGS;][]{Massey2006} data to make preliminary source classifications. They classified $\sim$ 200 sources as either foreground stars, ``stellar" sources, background galaxies/QSOs/AGN, XRBs, SNRs, and ``non-stellar" sources. Of these, 14 were classified as XRBs.

\subsection{{\it Spitzer} Catalog}\label{spitzer}

\begin{figure}
\centering
\includegraphics[width=0.5\textwidth,trim=20 0 0 0, clip]{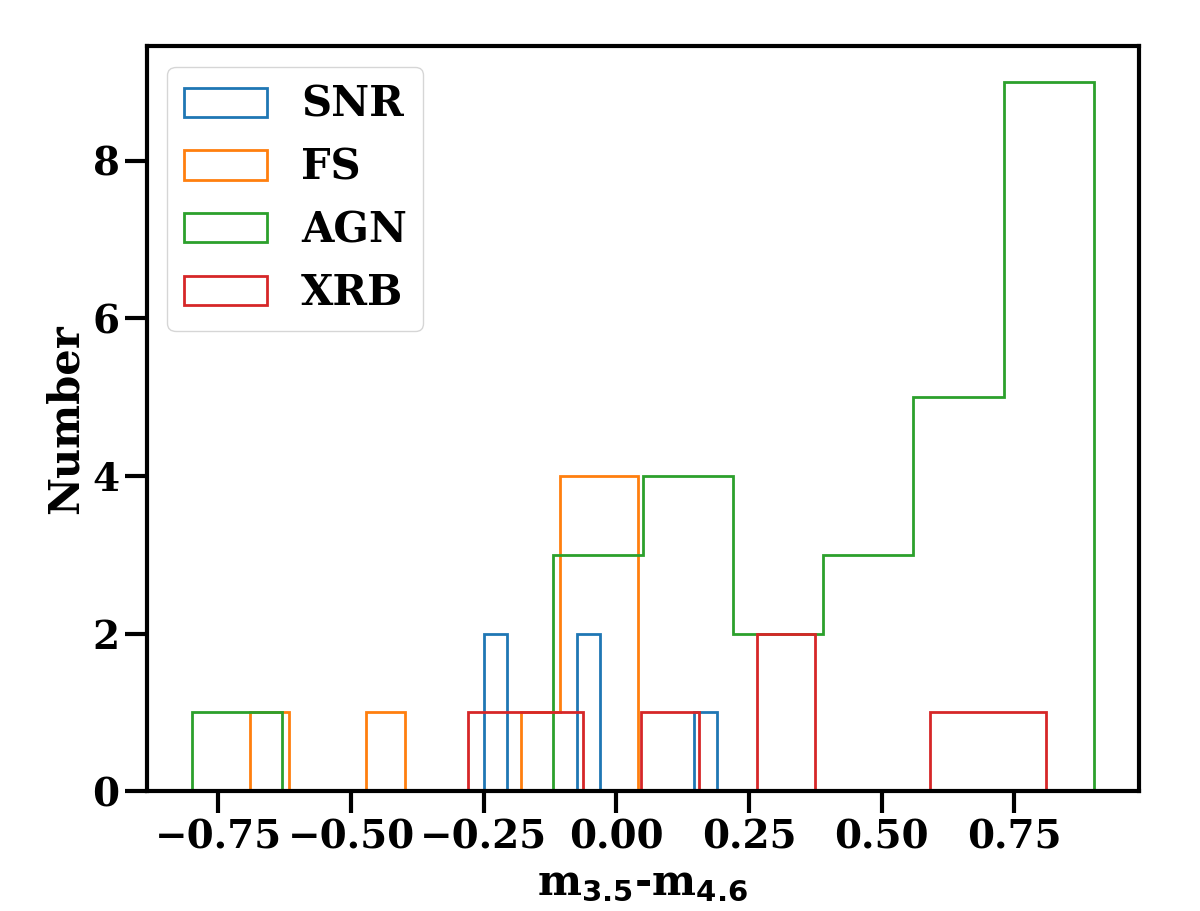}
\caption{Histogram of {\it Spitzer} $\mathrm{m_{3.5\mu m}-m_{4.6\mu m}}$ colors for X-ray sources from T11 with classifications as SNRs (blue), foreground stars (orange), XRBs (red), or AGN (green). The AGN generally have a larger color excess than other sources.} \label{spitzerhist}
\end{figure}

We cross-correlated all X-ray sources from T11 with the {\it Spitzer} catalog of \citet{Khan2015} to leverage the mid-IR colors of sources where available. As shown in \citet{Khan2015} galaxies display a color excess in $\mathrm{m_{3.5\mu m}-m_{4.6\mu m}}$, while sources like foreground stars fall along a nexus of zero color excess in these same bands. We used {\it Spitzer} colors, where available, as a coarse cut for separating X-ray sources that are most likely background galaxies, from other sources of interest (e.g. ``stellar" type sources). 

Sources with {\it Spitzer} colors $\leq$ 0 are most likely to be foreground stars and SNRs, while sources with {\it Spitzer} color excess (e.g. $>$ 0) are more likely to be background galaxies. We tested the efficacy of using {\it Spitzer} colors to separate X-ray sources of different types by first cross-correlating all {\it Spitzer} sources from \citet{Khan2015} in M33 with all T11 sources. We found 172 unique matches within search radii of 0.5". We then separated the sources based on their T11 classifications, as described in Section~\ref{chandra}, and binned the sources according to {\it Spitzer} colors as shown in Figure~\ref{spitzerhist}. 

We found that X-ray sources with the classifications of galaxy/QSO/AGN from T11 had {\it Spitzer} colors that were systematically higher than sources of other classifications. We therefore made a {\it Spitzer} color cut of $\mathrm{m_{3.5\mu m}-m_{4.6\mu m}}$ $>$ 0.5 to select X-ray sources that are likely background galaxies/AGN. This AGN color selection criteria may contain a small fraction of XRBs, but we combine the {\it Spitzer} color cuts with other criteria as described in Section~\ref{classify} to avoid excluding possible XRBs. Sources that are classified as SNRs or foreground stars in T11 unsurprisingly cluster around $\mathrm{m_{3.5\mu m}-m_{4.6\mu m}}$ $\leq$ 0, as shown in Figure~\ref{spitzerhist}. By contrast, sources that are classified as ``stellar", ``non-stellar," and XRB in T11 are spread across a range of {\it Spitzer} colors in this sample. Thus {\it Spitzer} colors, where available, are useful in narrowing down the set of X-ray sources that may be candidate HMXBs.  

\begin{figure}
\centering
\includegraphics[width=0.6\textwidth,trim=150 0 0 0, clip]{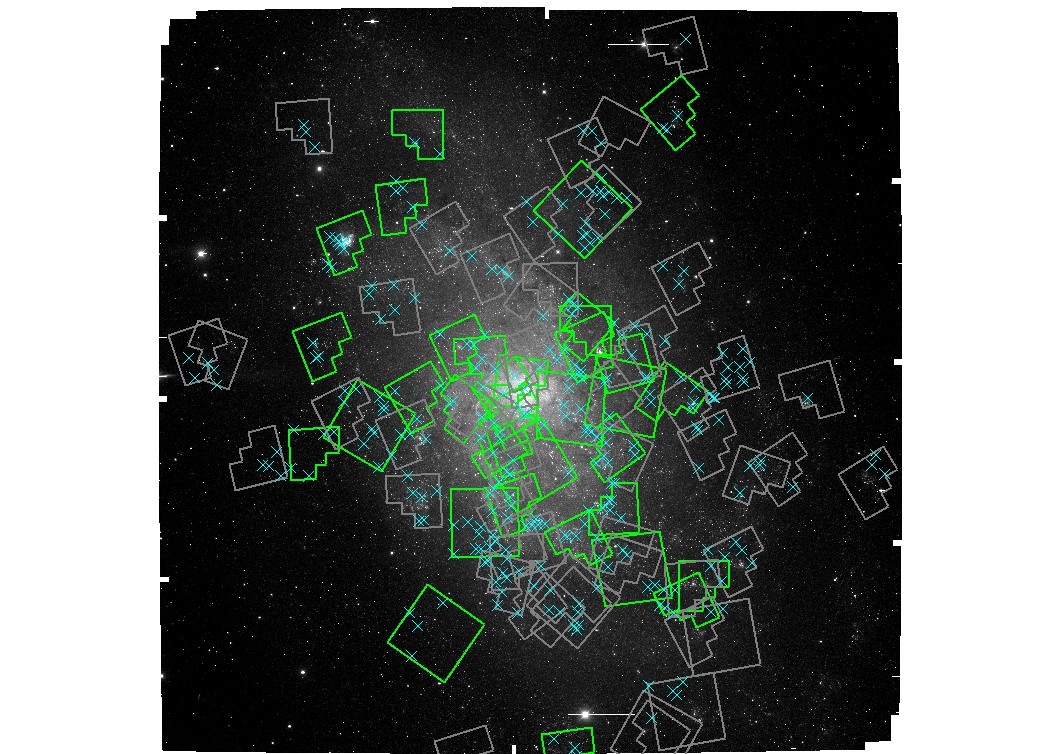}
\caption{R band Local Group Galaxy Survey \citep{Massey2006} image of M33 overlaid with all 89 archival {\it HST} fields (grey and green), and all 36 fields used in this analysis (green only). Cyan crosses are X-ray sources from \citet{Tullmann2011} that fall within the current archival {\it HST} coverage. There are 270 X-ray sources covered by 89 {\it HST} fields.} \label{m33coverage}
\end{figure}

\subsection{{\it HST} Photometry}\label{hst}

To search for optical counterparts to the X-ray sources in T11 we used all available archival {\it HST} data in M33 with coverage in at least two broad-band filters. This amounted to 89 fields covering about $\sim$ 40\% of the T11 X-ray catalog (270 unique sources). The full archival {\it HST} coverage of M33 (grey and green regions) and its overlap with the X-ray point source catalog (cyan crosses) is shown in Figure~\ref{m33coverage}. We reduced the {\it HST} data using the photometry pipeline developed as part of the ACS Nearby Galaxy Treasury Program \citep{Dalcanton2009} and the PHAT Program \citep{Dalcanton2012}, measured photometry using DOLPHOT and HSTPHOT \citep{Dolphin2000}, and ran artificial star tests as described in \citet{Williams2014}. All optical magnitudes reported here are Vega magnitudes.

For the artificial star tests, stars of known color and magnitude were inserted into the images, and the photometry was repeated to compare the output and input photometry. The resulting comparisons allow us to accurately model the photometric bias, uncertainty, and completeness as a function of color and magnitude. We used the photometry to construct CMDs of the stars within 50~pc of sources of interest (Section~\ref{candidates}), and artificial stars within 250~pc of sources in order to account for completeness and photometric uncertainties. Because the artificial star tests are performed on the whole field we choose this larger 250~pc region for the artificial stars to ensure that we have a large enough sample to quantify completeness.

We aligned all {\it HST} fields to a common frame as the X-ray data as described in Section~\ref{align}, which allows us to search for optical counterparts within the X-ray error circles for all sources with overlapping {\it Chandra} and {\it HST} coverage. Although we inspected all 89 archival fields for optical counterparts to X-ray sources to look for candidate HMXBs as described in Section~\ref{classify}, we ultimately present here only those fields in which there are HMXB candidates, which amounts to a subset of 36 of the 89 fields. This subset of 36 is displayed as green regions in Figure~\ref{m33coverage}. For reference, we list in Table~\ref{fieldtab} the proposal ID, field name, camera, filter set, exposure time, 50\% completeness magnitudes in both filters, and density of OB stars for all 36 {\it HST} fields used in this analysis.

\begin{table*}
\centering
\tiny
 \begin{threeparttable}
\caption{All archival {\it HST} fields used in this analysis. Column 1 lists the proposal ID number, column 2 is the field name, column 3 lists the camera, columns 4-5 are the {\it HST} filters available, columns 6-7 are the exposure times in seconds for each filter, columns 8-9 are the 50\% completeness limits in each filter for each field, and column 10 is the density of OB stars in each field.} \label{fieldtab}
\begin{tabular}{cccccccccc}
\hline \hline
Proposal ID & Name & Camera & Filter$_{1}$ & Filter$_{2}$ & Exposure$_{1}$ (s) & Exposure$_{2}$ (s) & m$_{1}^{50\%}$ & m$_{2}^{50\%}$ & OB star density (arcsec$^{-2}$) \\
\hline
10190 & M33-DISK1 & ACS & F606W & F814W & 2480 & 2480 & 25.80 & 25.10 & 0.09 \\
10190 & M33-146sw-26505 & ACS & F606W & F814W & 2160 & 2160 & 26.20 & 25.60 & 0.27 \\
10190 & M33-301sw-25900 & ACS & F606W & F814W & 2400 & 2500 & 26.10 & 25.40 & 0.19 \\
10190 & M33-626sw-28004 & ACS & F606W & F814W & 2160 & 2160 & 26.96 & 26.31 & 0.05 \\
10190 & M33-784se-549 & ACS & F606W & F814W & 2160 & 2160 & 27.60 & 26.80 & 0.04 \\
9873 & M33-267nw-31659 & ACS & F606W & F814W & 8800 & 17600 & 25.27 & 25.51 & 0.17 \\
10190 & M33-419se-160 & ACS & F606W & F814W & 2400 & 2500 & 27.48 & 26.81 & 0.03 \\
9873 & NGC598-U49 & ACS & F606W & F814W & 10414 & 20828 & 27.21 & 26.37 & 0.21 \\
9873 & NGC598-M9 & ACS & F606W & F814W & 10414 & 20828 & 28.21 & 27.37 & 0.12 \\
5998 & M33-FLD4 & WFPC2 & F336W & F439W & 800 & 800 & 23.23 & 23.90 & 0.05 \\
6640 & NGC598-SRV2$^{\bf s}$ & WFPC2 & F555W & F814W & 2240 & 2080 & 24.33 & 23.38 & 0.06 \\
5464 & NGC598-1$^{\bf s}$ & WFPC2 & F555W & F814W & 1600 & 1200 & 24.43 & 23.49 & 0.13 \\
5237 & M33-FIELDN604 & WFPC2 & F555W & F814W & 400 & 400 & 24.33 & 24.21 & 0.04 \\
5384 & M33-NEB1$^{\bf s}$ & WFPC2 & F336W & F439W & 320 & 360 & 22.22 & 22.85 & 0.08 \\
5998 & M33-FLD3 & WFPC2 & F336W & F439W & 800 & 800 & 23.10 & 23.18 & 0.12 \\
6640 & NGC598-SRV3 & WFPC2 & F555W & F814W & 2240 & 2080 & 24.98 & 24.13 & 0.18 \\
8207 & M33-PAR-FIELD7 & WFPC2 & F336W & F439W & 800 & 600 & 22.71 & 23.51 & 0.14 \\
6640 & NGC598-SRV5 & WFPC2 & F555W & F814W & 2240 & 2080 & 25.12 & 24.22 & 0.11 \\
11079 & M33-OB127 & WFPC2 & F336W & F439W & 600 & 490 & 21.81 & 23.80 & 0.16 \\
8207 & M33-PAR-FIELD2 & WFPC2 & F336W & F439W & 800 & 600 & 22.59 & 23.36 & 0.11 \\
8207 & M33-PAR-FIELD4 & WFPC2 & F336W & F439W & 800 & 600 & 22.75 & 23.36 & 0.08 \\
11079 & M33-OB137 & WFPC2 & F336W & F439W & 300 & 260 & 21.81 & 23.74 & 0.05 \\
11079 & M33-OB90 & WFPC2 & F336W & F439W & 600 & 490 & 21.62 & 23.80 & 0.18 \\
6431 & NGC598-FIELD & WFPC2 & F555W & F814W & 520 & 460 & 25.08 & 24.10 & 0.02 \\
5914 & NGC598-R12 & WFPC2 & F555W & F814W & 4800 & 5200 & 25.31 & 24.67 & 0.12 \\
5914 & NGC598-R14$^{\bf s}$ & WFPC2 & F555W & F814W & 4800 & 5200 & 24.75 & 23.92 & 0.13 \\
5914 & NGC598-U137 & WFPC2 & F555W & F814W & 4800 & 5200 & 26.25 & 25.62 & 0.11 \\
6038 & M33-AM6-FIELD & WFPC2 & F336W & F439W & 1800 & 600 & 23.64 & 23.37 & 0.07 \\
8018 & M33 & WFPC2 & F555W & F814W & 3800 & 3800 & 25.60 & 24.80 & 0.10 \\
11079 & M33-OB39S$^{\bf s}$ & WFPC2 & F336W & F439W & 600 & 490 & 21.77 & 22.37 & 0.17 \\
11079 & M33-OB77$^{\bf s}$ & WFPC2 & F336W & F439W & 600 & 490 & 21.73 & 22.46 & 0.12 \\
11079 & M33-OB94$^{\bf s}$ & WFPC2 & F336W & F439W & 600 & 490 & 21.76 & 22.33 & 0.22 \\
11079 & M33-OB101$^{\bf s}$ & WFPC2 & F336W & F439W & 600 & 490 & 21.65 & 22.39 & 0.20 \\
9127 & M33-PAR-FLD1$^{\bf s}$ & WFPC2 & F336W & F439W & 520 & 600 & 22.14 & 22.86 & 0.12 \\
5384 & M33-NEB6$^{\bf s}$ & WFPC2 & F336W & F439W & 320 & 360 & 22.40 & 23.00 & 0.22 \\
5494 & M33-OB6-5-FIELD$^{\bf s}$ & WFPC2 & F300W & F555W & 350 & 300 & 22.44 & 25.19 & 0.20 \\
\hline
\end{tabular}
\begin{tablenotes}
      \small
      \item [s] Field depth does not reach 80 Myr MSTO. 
    \end{tablenotes}
  \end{threeparttable}
\end{table*}

\subsection{Image Alignment}\label{align}

Robustly identifying optical counterparts to X-ray sources requires precise astrometric alignment between the archival {\it HST} data and the pre-existing {\it Chandra} catalog from T11. We chose 2MASS as a common frame for the alignment procedure, as the source positions in the T11 catalog are already aligned to 2MASS. 

We used the R-band photometry from the LGGS \citep{Massey2006} with source positions updated as prescribed in \citet{Massey2016} as the standard reference for aligning all {\it HST} fields. This choice was made to ensure that there were enough stars in each {\it HST} field  with counterparts in the reference image on which to align. To use the R-band photometry from the LGGS as the standard reference for aligning each {\it HST} field, we first had to make sure that the LGGS positions were also aligned to 2MASS. To do so we chose $\sim$ 1000 sources from the R-band photometry between 18th and 20th magnitude and found each source's corresponding reference source in 2MASS. We then aligned the two images using the {\tt PyRAF} task {\tt CCMAP} with 2MASS as the reference. The $\Delta$RA and $\Delta$Dec fit RMS values from this procedure were 0.19", and 0.16", respectively. 

We next aligned each archival {\it HST} field to the pre-aligned R-band image, which results in {\it HST} fields that are aligned to 2MASS, and thus a common frame as the X-ray data from T11. We chose stars between 18th and 22nd magnitude in each {\it HST} field, with the exact magnitude limits depending on the depth of the field. In general, we set the limits so that we found at least three R-band counterparts to the {\it HST} stars on which to align. We again aligned the images using the {\tt PyRAF} task {\tt CCMAP} with the 2MASS aligned R-band image as the new reference frame for each {\it HST} field. We performed this procedure in an iterative fashion, so outlier source matches were culled from the {\tt CCMAP} input, and the task repeated prior to computing the final plate solution. 

We then assessed by eye the alignment between each archival field and the R-band image to ensure good agreement between source centroids in the images, as well as the distribution of separations for {\it HST} sources and their reference counterparts in the R-band image. In general, the  $\Delta$RA and $\Delta$Dec RMS fit values from this process were $<$ 0.1", and often smaller. We added in quadrature the $\Delta$RA and $\Delta$Dec fit RMS from the Rband to 2MASS and {\it HST} to R band alignment procedures, which makes the positional error due to the full alignment process $\sim$ 0.2" for most sources. This alignment error is smaller than the X-ray source position error itself. The X-ray positional error for each source comes directly from the T11 catalog, which has an error floor of 0.5", though the position errors may be larger depending on off-axis angle and source intensity. The total error due to the alignment procedure ($\sim$ 0.2") is then added to the {\it Chandra} position error from T11 ($\sim$ 0.5" for the majority of sources considered here, but sometimes higher for sources that are far off-axis), so the final X-ray position errors are a combination of both intrinsic source position uncertainty, and uncertainty from the alignment process. This results in X-ray error circles that are typically $<$ 0.7" in radius for our sources. With this kind of precision astrometric alignment is possible to search for optical counterparts to the X-ray sources from T11. 

\subsection{Identifying Counterpart Candidates}\label{classify}

After aligning all available {\it HST} images to a common frame with the T11 catalog, and cross-correlating the \citet{Khan2015} {\it Spitzer} catalog with the X-ray source catalog we leveraged the multi-wavelength coverage to identify candidate HMXBs in M33. We separated HMXBs from other X-ray bright sources, including foreground stars, SNRs, and AGN based on a combination of X-ray HRs, {\it Spitzer} colors (if available), magnitude and color of the {\it HST} source or sources within the X-ray error circle, X-ray variability, and classification based on visual inspection of the {\it HST} images (e.g. for extended sources) as described in turn below. 

We first made cuts on X-ray HRs as defined in Figure~\ref{hr_box} to separate sources broadly into the categories of ``soft" (SNRs and foreground stars), ``intermediate" (AGN, HMXBs, and LMXBs), and ``hard" (potentially AGN and HMXBs behind large column densities). For sources with {\it Spitzer} counterparts (65 sources) the categorization as SNR or foreground star was strengthened if the {\it Spitzer} colors were $\leq$ 0. Foreground stars can sometimes be identified by visual inspection, and SNRs were likewise easily identified via cross-correlation with the SNR catalogs of \citet{Long2010} and \citet{Garofali2017}. 

If the {\it Spitzer} color was in excess of 0.5 we considered the source a strong AGN candidate, while if the {\it Spitzer} colors were between 0--0.5, or if there were no {\it Spitzer} counterparts available we relied on further measures, such as variability, optical counterpart color, and the aforementioned HRs to further classify the source as either an AGN or XRB candidate. The efficacy of the {\it Spitzer} color cuts is demonstrated in Figure~\ref{spitzerhist} as compared to previous source classifications from T11. We considered detection of short or long term variability from T11 to be indicative of a possible HMXB, though AGN may also be variable on such timescales \citep{Pao2004}. 

\begin{figure}
\centering
\includegraphics[width=0.5\textwidth,trim=20 0 0 0, clip]{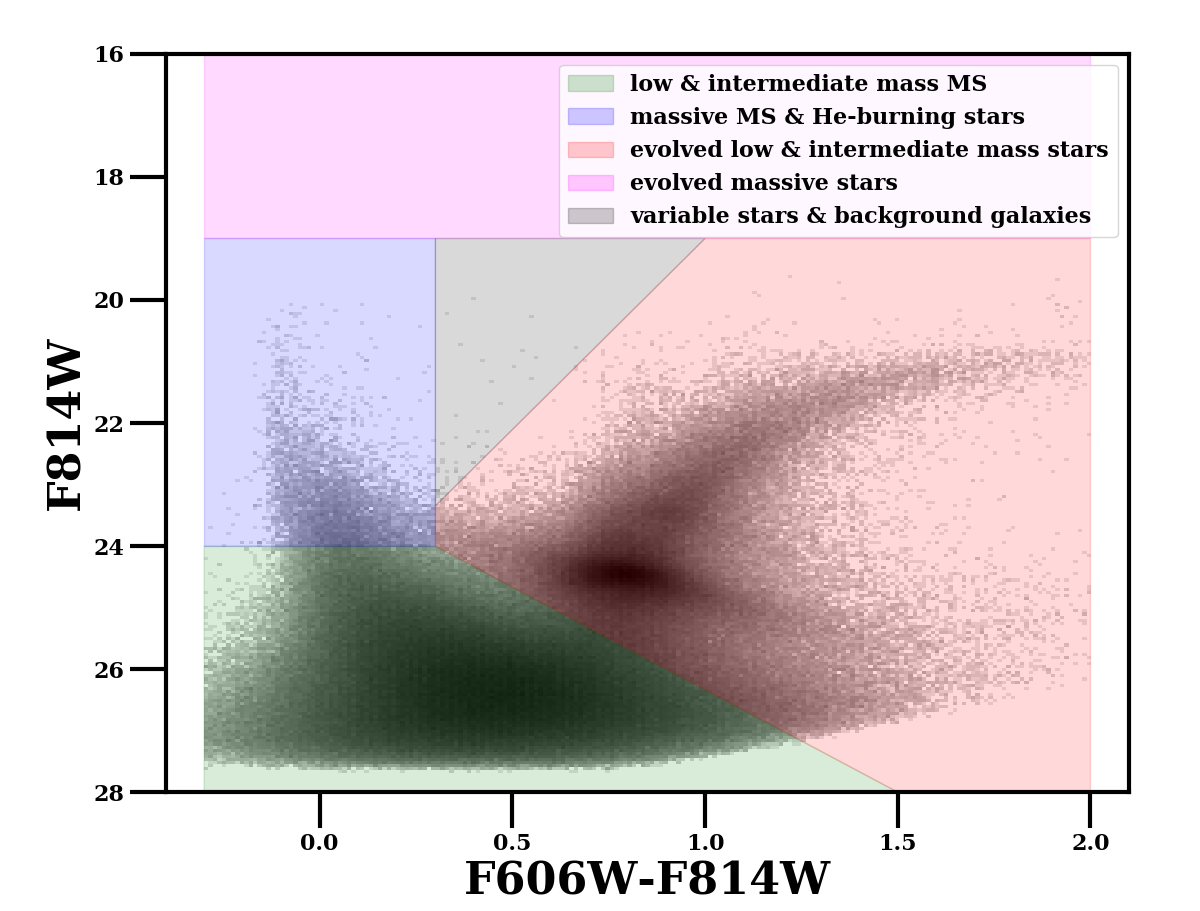}
\caption{A Hess diagram constructed using all the stars in one of the {\it HST} ACS fields binned by 0.05 magnitudes in both color and magnitude space. The shaded regions delineate sources in different mass ranges and stages of stellar evolution: green: low and intermediate mass main sequence stars; blue: massive main sequence stars (e.g. OB stars) and He-burning stars; red: evolved low and intermediate mass stars (e.g. red giant branch, red clump, and asymptotic giant branch stars); magenta: evolved massive stars (e.g. supergiants); grey: variable stars and background galaxies. Optical counterparts for HMXBs would lie in the blue and magenta shaded regions.} \label{annotatedcmd}
\end{figure}

XRBs (both LMXBs and HMXBs) and AGN are not easily separated from one another based on one measure alone. All such sources fall within the same general area of the X-ray HR plot (black box, Figure~\ref{hr_box}), and may both be time variable, but should be associated with markedly different optical counterparts in terms of their spectroscopic and photometric characteristics. As Hectospec data is only available for a small subset of the T11 catalog we rely primarily on a photometric classification scheme, using the complementary Hectospec spectra where available. In this way, the {\it HST} data are crucial for separating HMXBs, which should have relatively bright, blue stars as their optical counterparts, from background AGN and LMXBs, whose counterparts should appear much redder, and in the case of AGN, sometimes more extended. In Figure~\ref{annotatedcmd} we show the loci for optical sources in different mass ranges and stages of stellar evolution as shaded regions overlaid on a Hess diagram constructed from one of the ACS fields used in this work. An optical counterpart in the blue or magenta regions of the Hess diagram in Figure~\ref{annotatedcmd} that falls within the error circle of an X-ray source makes that X-ray source a strong HMXB candidate.

With all X-ray and {\it HST} data aligned to a common frame we constructed CMDs for regions within 50~pc of all X-ray sources, and noted the positions on the CMD of any {\it HST} sources that were within the X-ray error circle, where the error circle is derived from a combination of the individual source positional uncertainties from T11, which account for off-axis angle and source intensity, and the uncertainty due to alignment, as described in Section~\ref{align}. Because we are interested primarily in good HMXB candidates we selected optical counterparts within the X-ray error circle that were brighter than 24th magnitude (roughly the magnitude of a B star at the distance of M33), and had F606W-F814W, or F555W-F814W colors $\leq$ 0.3 (bluer than the red giant branch). Assuming a Milky Way like extinction curve (i.e. R$_{v}$ = 3.1), this color cut allows for about one magnitude of extinction for stars on the main sequence.

\begin{figure*}
\begin{minipage}[b]{.33\linewidth}
\centering%
\includegraphics[width=\linewidth,trim=20 0 20 0, clip]{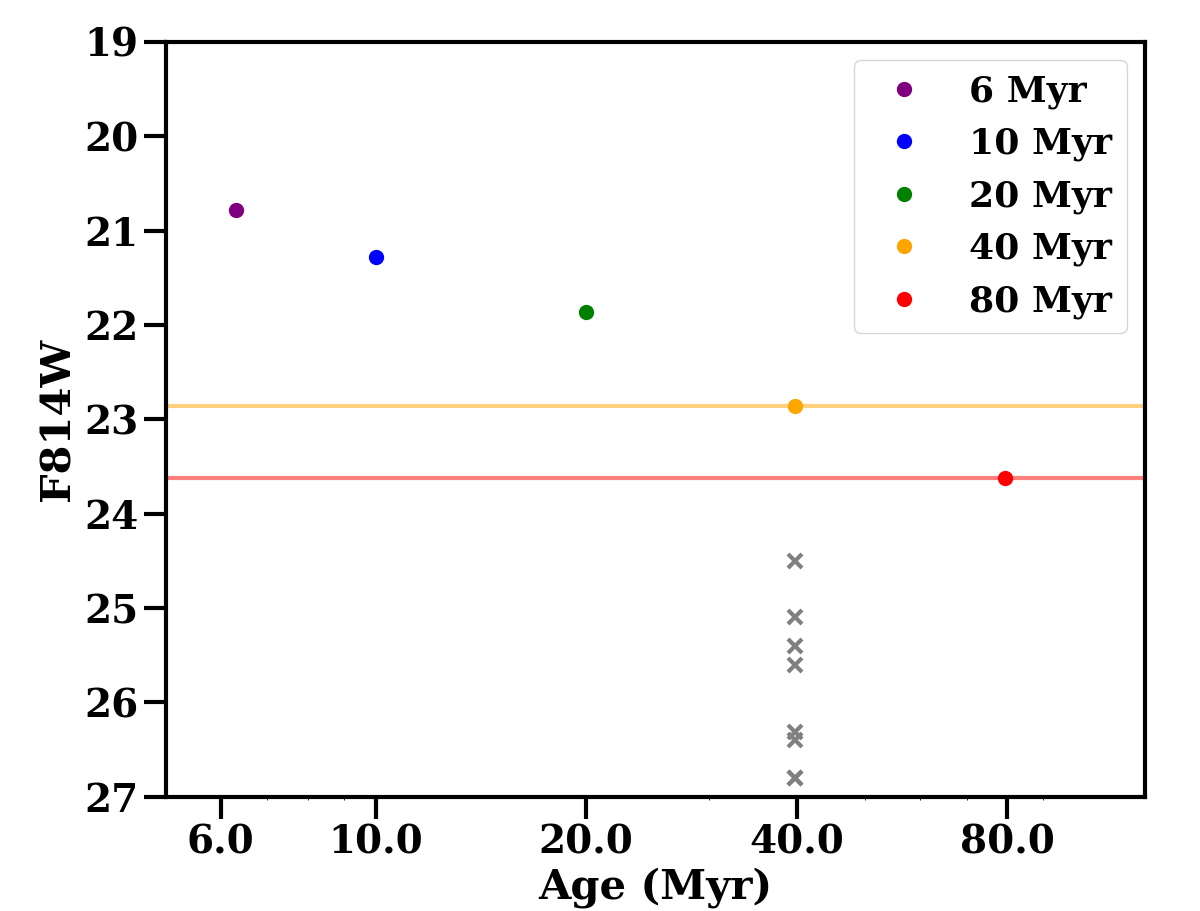}
\end{minipage}%
\hfill%
\begin{minipage}[b]{.33\linewidth}
\centering%
\includegraphics[width=\linewidth,trim=20 0 20 0, clip]{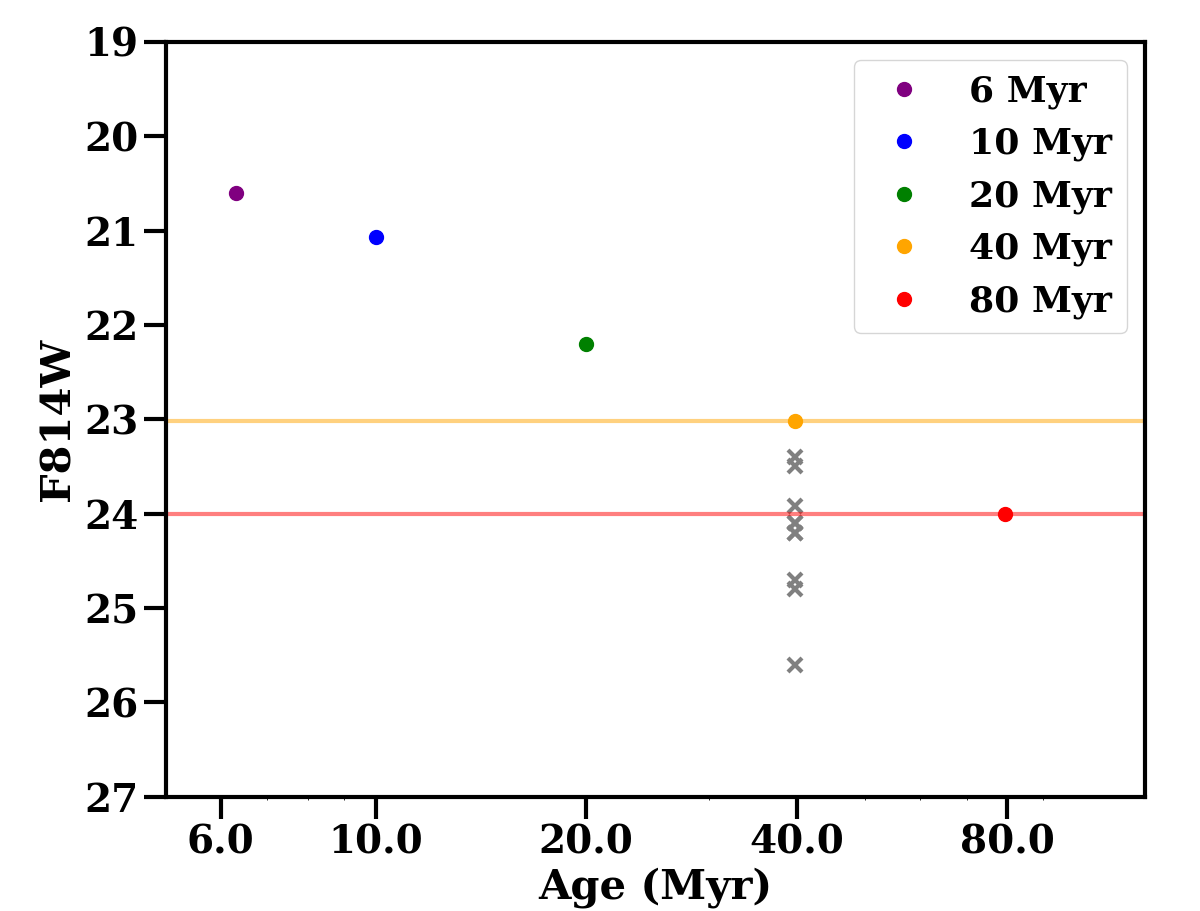}
\end{minipage}
\hfill%
\begin{minipage}[b]{.33\linewidth}
\centering%
\includegraphics[width=\linewidth,trim=20 0 20 0, clip]{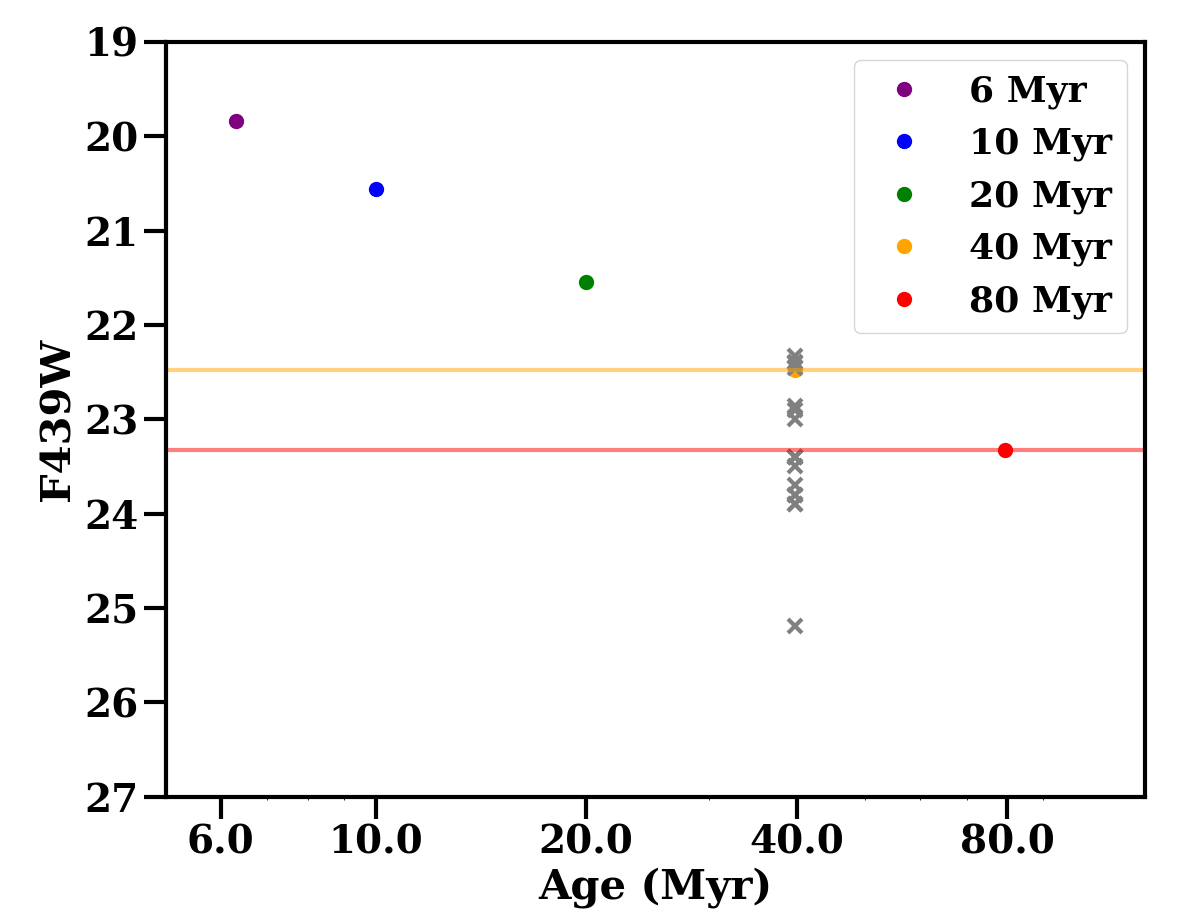}
\end{minipage}
\hfill%
\caption{{\it Left}: The 50\% completeness magnitudes in the F814W filter for all ACS fields (grey points) at a MSTO of $\sim$ 40 Myr. The MSTO magnitudes for ages from 6-80 Myr are plotted in color and labeled for reference. The 50\% completeness limits for all ACS fields are well above the MSTO magnitude at 80 Myr, indicating that we are sensitive the MSTO at 80 Myr in these fields. {\it Center}: The 50\% completeness magnitudes in the F814W filter for all WFPC2 fields (grey points) at a MSTO of $\sim$ 40 Myr. Three fields (6640-NGC598-SRV2, 5464-NGC598-1, 5914-NGC598-R14) have 50\% completeness magnitude limits fainter than the MSTO magnitude at 80 Myr, indicating that we are not sensitive to the MSTO of the oldest populations in these fields. {\it Right}: The 50\% completeness magnitudes in the F439W filter for all WPC2 fields (grey points) at a MSTO of $\sim$ 40 Myr. There are eight fields (5384-M33-NEB1, 11079-M33-OB39S, 11079-M33-OB77, 11079-M33-OB94, 11079-M33-OB101, 9127-M33-PAR-FLD1, 5384-M33-NEB6, and 5494-M33-OB6-5-FIELD) with a completeness limits fainter than the MSTO magnitude at 80 Myr, indicating that these fields are not sensitive to the MSTO of the oldest populations}\label{completeness}
\end{figure*}

Identification of a bright, blue star within the X-ray error circle does not denote a unique optical counterpart to the X-ray source, however because bright OB stars have low surface density the presence of such a star within the X-ray error circle suggests the source is a good candidate HMXB. We estimate the chance coincidence probability, and thus the number of possible spurious sources by first calculating the density of bright, blue stars in each field by selecting stars brighter than 24th magnitude and bluer than F606W-F814W, or F555W-F814W = 0.3. The density of OB stars for each field is listed in the last column of Table~\ref{fieldtab}. These densities taken in conjunction with a typical X-ray error circle of 0.7" yield the chance coincidence probability for each field. To provide a representative value for the whole sample, we calculate a weighted average of the chance coincidence probabilities in each field, weighted by the number of candidate HMXBs in that field. Based on this calculation we expect 0.21 bright, blue stars are chance coincidences with the typical X-ray error circle. This implies that of our 55 candidate HMXBs 11 $\pm$ 3 may be chance coincidences of blue stars with the X-ray source position, and thus spurious HMXB candidates. Further discussion of potential spurious HMXB candidates in our sample, and their effects on our results are presented in Section~\ref{spurious}. 

The above color and magnitude cuts will select relatively bright, blue sources, but will miss any bright sources that are slightly more red in color. Sources much redder than these color cuts that fall within our X-ray error circles may either be background AGN, evolved massive stars, or otherwise potentially stars with strong emission lines (e.g. emission line stars with strong H$\alpha$). For this reason, we also included as candidate HMXBs sources that had optical counterparts in their error circles that were brighter than $\sim$ 21st magnitude, but nominally more red than F606W-F814W or F555W-F814W = 0.3, though not as red as the red giant branch stars. Again assuming R$_{v}$ = 3.1 this more lenient color cut allows for about two magnitudes of extinction for main sequence stars. We then checked special cases for any further classification from T11 on the basis of available Hectospec spectra. If the X-ray source had an {\it HST} counterpart brighter than $\sim$ 22nd magnitude, and corresponding Hectospec data with classification ``stellar" in T11 we include it as an HMXB candidate. If the source was classified as ``QSO/AGN" in T11 on the basis of available optical spectroscopy then it was excluded. Sources that were classified as ``XRB" in T11 were also automatically included, even if they did not pass the above color cuts. Finally, to determine our list of candidate HMXBs all sources were inspected visually, and compared against the classification determined from the color cuts at X-ray, optical, infrared wavelengths, and previous classifications available from T11. Notes on individual sources are provided in Table~\ref{candidate_tab}.

\subsection{Age Constraints from Surrounding Populations}\label{cmdfit}

We measure ages, or formation timescales for the HMXB candidates in this sample using the stars surrounding each candidate. This process involves first extracting photometry of stars in the vicinity of HMXB candidates, and then fitting the photometry using stellar evolution models. We detail both of these steps below. 

\subsubsection{Selecting Stellar Samples}\label{sizeselect}

For young sources it should be possible to determine the age, or formation timescale of the source based on the surrounding stellar population. Most stars are known to form in clusters, and thus in relatively co-eval populations that stay associated in regions of $\sim$50 pc on timescales $<$ 60 Myr \citep{Lada2003,Gogarten2009,Eldridge2011}.  In the case of HMXBs, the system age is limited to be less than the lifetime of the secondary, which is a high-mass star with a lifetime of $\lesssim$50~Myr.  Thus, we can potentially determine ages for HMXBs by fitting the CMDs of the stellar population within 50~pc of the source. This measured age is the {\it total} age of the binary, which is a combination of the time to formation for the compact object from the primary, the delay time between compact object formation and the onset of the X-ray emission, and the time since the start of X-ray emission.

Such measurements of the ages of young objects have been performed successfully to constrain the ages and masses of the progenitors of supernovae and supernova remnants \citep[e.g.][]{Jennings2012,Jennings2014,Williams2018}, which have similar maximum ages (and therefore associated region sizes) as HMXBs as set by the lifetimes of massive stars.  We therefore adopt this same methodology for attempting to constrain the ages of HMXB candidates.

While the 50~pc region is commonly-used for constraining ages of young populations, because the HMXB phase may occur millions of years after the formation of the compact object, there is some possibility that the HMXB could have moved farther from its associated young population following compact object formation. Namely, there is some possibility that the binary could have received a significant kick during compact object formation.  While the nature and strength of such kicks is highly uncertain \citep{Lipunov1997,Pod2004,Janka2012}, such a kick could potentially send an HMXB into isolation, or, more problematically for this kind of analysis, into a neighboring star-forming region which contains a young stellar population not representative of the true HMXB age. This scenario is possible, but theory and observations suggest that large natal kicks and long delays between compact object formation and the onset of X-ray emission may not be common \citep[e.g.][]{Pfahl2002,Mirabel2003,Linden2009,Linden2010,Repetto2015,Mandel2016}. The offsets between HMXBs and nearby young clusters in the SMC are $\sim$65~pc \citep{Coe2005}, roughly consistent with a 50~pc association region. Similarly, \citet{Sepinsky2005} find median distances between XRBs and their natal clusters of 30-100~pc for 15-50 Myr age clusters in binary population synthesis simulations that assume simple cluster models. For potential cases of very large kicks and long travel times,  we would expect to see at least some sources kicked into regions without any young stars (i.e. no recent star formation), but we do not find evidence of this in our M33 sample.

We test that our results are not highly dependent on our choice of extraction region size by extracting and fitting samples from two different region sizes for each HMXB candidate.  The first sample corresponds to the 50~pc region typically used in studies of young population ages, while the second is double this radius. We find that the results are consistent in both cases, and so primarily discuss the results using the 50~pc regions in Section~\ref{sec:agedist}.

\subsubsection{Color-Magnitude Diagram Fitting}\label{technique}

In order to measure resolved SFHs and thus recover ages for HMXB candidates we modeled the CMDs of the stellar population within 50~pc of the X-ray source of interest using the software MATCH \citep{Dolphin2002}. MATCH fits the observed CMD by building up synthetic diagrams using the stellar evolution models of \citet{Girardi2002} and \citet{Marigo2008} for a range of ages and metallicities and finding the combination that best fits the observed distribution. For this analysis we assumed a distance modulus of 24.67 for M33 \citep{Rizzi2007}, a Kroupa initial mass function (IMF), and a binary fraction of 0.35. For sources within 3 kpc of the nucleus we modeled the CMDs over a range of metallicities from [Fe/H] = -0.4 to 0.00 with steps of 0.1 in log Z, while for sources outside 3 kpc we chose a range of [Fe/H] = -0.6 to -0.3 in accordance with the measured metallicity gradient in M33 \citep{Magrini2007}. We used bin sizes of 0.2 in magnitude space, and 0.1 in color space for binning the CMD. Our age bins were of width 0.1 dex and range from 4.0 Myr--14.1 Gyr. MATCH places any star formation $<$ 5 Myr in the 4-5 Myr time bin, therefore age measurements in the youngest time bins are actually upper limits on the age (e.g. $<$ 5 Myr), and provide lower limits on the progenitor mass when mapping from age to mass. We fit for a range of extinction values, and included a prescription for differential reddening.

In fields with very shallow depth the CMD may not be sensitive to stellar populations at older ages, which means the CMD fitting routine will not be able to place strong constraints on the star formation in older age bins. For each field we used artificial stars to quantify the field depth as the 50\% completeness magnitude limit (Section~\ref{hst}). The completeness limits for each field are listed in columns 7-8 in Table~\ref{fieldtab}. To ensure that we are sensitive to the main sequence turn-off (MSTO) for various population ages we compared the 50\% completeness magnitudes in each field to the MSTO magnitudes at ages from 6-80 Myr  from the stellar evolution isochrones of \citet{Girardi2002} in the same filters used here. As shown in the left-hand panel of Figure~\ref{completeness} we are sensitive to MSTO at 80~Myr for all ACS fields. The center and right-hand panels of Figure~\ref{completeness} display the same comparison between 50\% completeness magnitude limits in each field (grey x's) and MSTO magnitudes at various ages in the WFPC2 filters used. Here, there are several fields where the depth is so shallow that we are not quite sensitive to the magnitudes of the oldest MSTO ages. We have flagged these fields (11) for which we are not sensitive to the MSTO at 80 Myr with $s$ in Table~\ref{fieldtab} and Table~\ref{candidate_tab}, and note that the age estimates for sources in these fields may have large errors toward older ages.  

\begin{figure*}
\begin{minipage}[b]{.4\linewidth}
\centering%
\includegraphics[width=\linewidth,trim=0 0 0 0, clip]{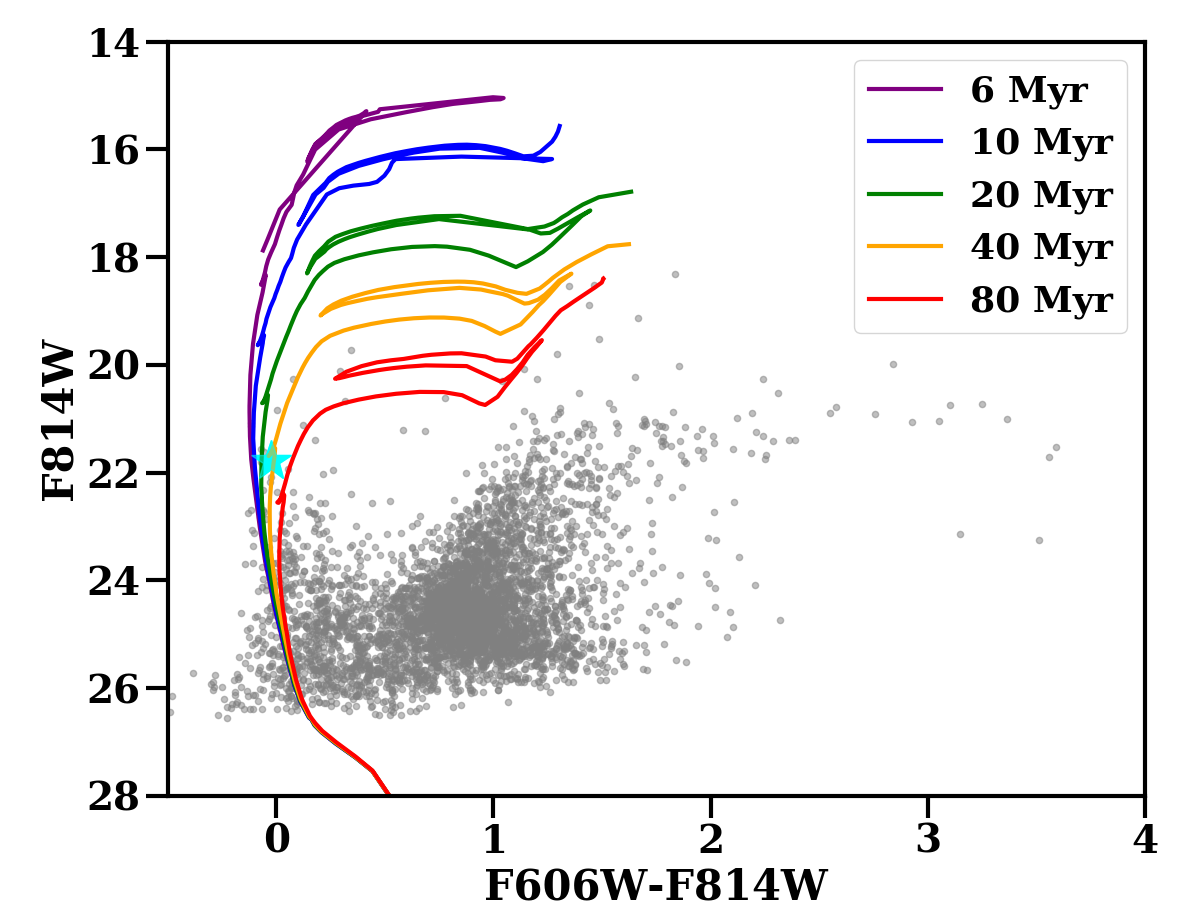}
\end{minipage}%
\hfill%
\begin{minipage}[b]{.6\linewidth}
\centering%
\includegraphics[width=\linewidth,trim=0 0 0 0, clip]{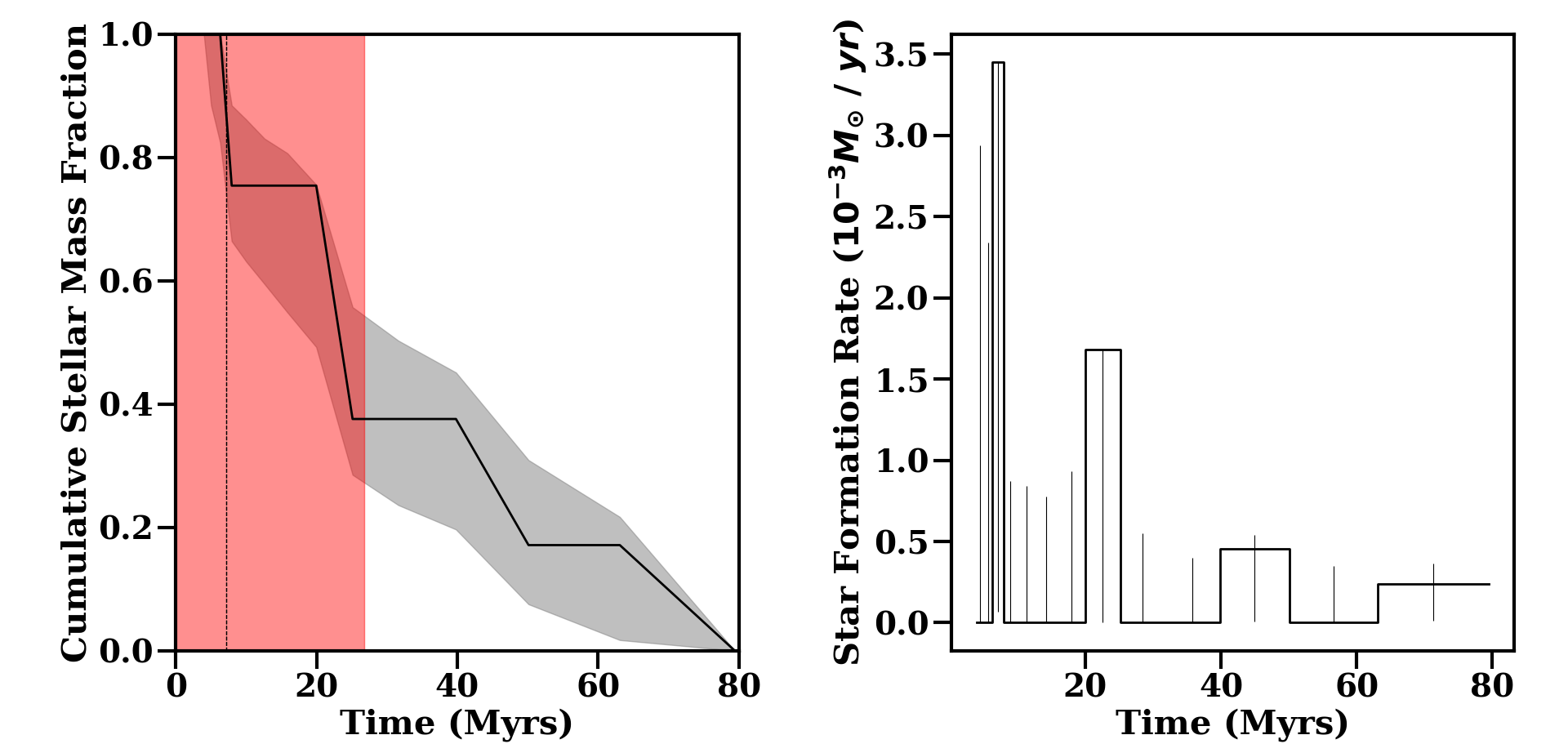}
\end{minipage}
\hfill%
\caption{{\it Left}: The CMD of stars within 50 pc of the HMXB candidate 013341.47+303815.9 (grey points), with the candidate optical counterpart denoted by the cyan star. Padova group isochrones shifted to the distance and extinction of M33 are overlaid for reference. {\it Center}: The cumulative stellar mass fraction (black line) formed over time based on modeling the CMD of the stars within 50 pc of the HMXB candidate. The Monte Carlo derived errors on this distribution are in grey. The most likely age is the bin with the highest fraction of stellar mass formed, and is denoted by the dotted line. The narrowest 68th percentile confidence interval on the most likely age are denoted in red. {\it Right}: The SFH for this candidate HMXB based on modeling the CMD, indicative of a young source.}
\label{cmdfit_ex}
\end{figure*}

In addition to quantifying errors based on photometric depth we quantified systematic errors by randomly resampling the CMD and refitting with model isochrones shifted in both bolometric magnitude as well as effective temperature \citep{Dolphin2012}. We quantified random errors from the fitting technique itself using MATCH's {\tt hybridMC} routine which uses a Markov Chain Monte Carlo approach to quantify uncertainties on the amount of star formation in a given age bin by sampling from the probability density distribution of the fit values \citep{Dolphin2013}. The random error from the fit method most often affects the relative amount of star formation in a particular age bin rather than the relative age bin localization of a star formation burst. 

The SFH returns the SFR in each time bin, and we use this to derive the cumulative distribution of stellar mass using the SFR at each time step to calculate first the total stellar mass formed in the timespan of interest, and then the fraction of this total stellar mass formed in each time bin. The cumulative distribution of stellar mass that is returned is akin to a probability distribution function with the most likely age for the source being the time bin with the highest fraction of stellar mass formed. Errors on the overall cumulative distribution are computed by resampling the errors on the SFRs in each time bin 1000 times using a Gaussian distribution with mean of the best SFR and sigma of the rate of uncertainty.

We report the most likely age for a given source as the time bin with the highest fraction of stellar mass formed. The errors on this age are given here as the narrowest 68th percentile confidence interval, calculated using 1000 realizations of the SFH from MATCH's {\tt hybridMC} routine \citep{Dolphin2013}. As described in \citet{Murphy2018}, the more common use of the 16th and 84th percentiles as the confidence interval on a measurement introduces biases in the cases where the probability distribution function has large ``wings." As this is sometimes the case for data such as this, the use of the narrowest 68th percentile tempers the effects that such biases would have on the reported age and associated errors.

Even in the case of multiple star formation events we are therefore able to measure the most likely age using the time bin with the highest fraction of stellar mass formed, and report the narrowest 68th percentile confidence interval using the full probability distribution function spanning the full 80~Myr. We determined the age for a source using only the last 80 Myr of the SFH as we are interested in young systems hosting neutron stars or black holes. Given that at least a $\sim$ 8 M$_{\odot}$ progenitor is needed to produce a neutron star \citep{Jennings2012,Jennings2014} all neutron stars should be formed within $\sim$ 50 Myr of the star formation episode. We therefore probed out to 80 Myr to allow for some delay of the onset of the HMXB phase given considerations of binary evolution, but note that we are not particularly sensitive to ages much older than this because dynamical mixing of populations means we are limited by photometric depth. Furthermore, contamination begins to drown out signal at older ages. 

 An example the cumulative stellar mass fraction and associated SFH for HMXB candidate 013341.47+303815.9 is shown in the center and right panels of Figure~\ref{cmdfit_ex}. The most likely age (dotted line) based on the time bin with the highest fraction of stellar mass formed in the last 80 Myr is 7.1~Myr (narrowest 68th percentile confidence interval in red). The CMD of stars within 50 pc of this source is shown on the left with stellar isochrones overlaid for reference. The most likely optical counterpart within the X-ray error circle is denoted by the cyan star, and is consistent with a young age based on the isochrones. 
 

\section{Results}\label{results}

We identified 55 candidate HMXBs in M33 based on their X-ray, optical, and infrared characteristics as described in Section~\ref{classify}. The majority ($\sim$ 78\%) were previously unclassified as potential XRBs. We measured an age for each source using the CMD fitting method described in Section~\ref{cmdfit}. Here we describe the sample of HMXB candidates that fall within our archival {\it HST} coverage, their optical counterparts, the age measurement results for interesting sources in M33, and the resulting combined age distribution for all HMXB candidates in M33 within the archival {\it HST} coverage. 

\subsection{Candidate HMXBs in M33}\label{candidates}

Based on the XLF measured by \citet{Williams2015} and T11, M33 is expected to host a large HMXB population. 14 sources in the T11 catalog (662 sources) were classified as possible XRBs. Our technique has allowed us to nearly triple the number of candidate HMXBs in M33.

We identified 55 HMXB candidates from the 270 X-ray sources in T11 for which there is overlapping {\it HST} coverage. Given the relation between the number of HMXBs and SFR from \citet{Grimm2003}, we infer that there are likely $\sim$ 109 HMXBs in M33 above a limiting luminosity of $\sim$ 1 $\times$ 10$^{35}$ ergs~s$^{-1}$, roughly consistent with the limiting luminosity at 90\% completeness in T11. Considering that the overlapping {\it HST} coverage accounts for only 40\% the T11 survey area we therefore expect to find around 43 HMXBs in our sample, consistent with the 55 sources with 11 $\pm$ 3 spurious candidates reported here.

Of the 55 candidate HMXBs reported here, eight were previously classified as possible XRBs by T11 (with the remaining six T11 XRBs outside this survey area). All candidate HMXBs from this study are listed in Table~\ref{candidate_tab} with corresponding T11 catalog identifier, {\it HST} field name, RA/Dec from T11, deprojected galactocentric radius in M33, source type from T11, match from \citet{Grimm2005} (described below), and source notes. 

Fields where the depth was not sufficient for our CMD fitting routine to be sensitive to the MSTO of stellar populations older than 80 Myr are denoted by $s$ in Table~\ref{candidate_tab} next to the field name. As noted in Section~\ref{technique} the age measurements for such sources will have larger errors at older ages, and thus are not as reliable as measurements from deeper fields. 

We compared our catalog of candidate HMXBs in M33 with those of \citet{Grimm2005,Grimm2007} to look for any previous classifications. The \citet{Grimm2005} catalog contains, as a subset, 16 unique X-ray sources with optical counterparts in M33, i.e. candidate HMXBs. Of these 16, all are in the T11 coverage area, but only 12 are also in our archival HST fields, so we matched only to these 12. We found matches for these sources to T11 positions by adopting a search radius of 4.5". Six of these 12 matches are also classified as candidate HMXBs in this work (noted in column seven of Table~\ref{candidate_tab}, candidate optical counterparts in Table~\ref{opt_tab}), while the remaining six sources from \citet{Grimm2005} are classified here as either background AGN (two sources), SNRs (three sources), or foreground stars (one source). 

The notes column in Table~\ref{candidate_tab} lists more detailed information on each source: ``no {\it HST} counterpart" if the source had no reliable {\it HST} optical counterpart either due to shallow exposures in available filters or issues of stellar crowding,  ``low SFR" if the source had SFR $<$ 4 $\times$ 10$^{-4}$ M$_{\odot}$ yr$^{-1}$ in all time bins, or $<$ 1000 M$_{\odot}$ formed in 80 Myr, ``no SFH" if there were too few stars in the field to recover a SFH,``red counterpart" if the source had a bright ($<$ 21st magnitude) counterpart in its error circle that did not pass our blue color cuts, or similarly a $<$ 22nd magnitude counterpart with previous classification as ``stellar", ``SNR?" if the source was previously classified as an SNR in T11 and \citet{Long2010}, or ``$>$ 2$\sigma$ var or $>$ 3$\sigma$ var" if the source showed variability between the deep observations of T11 and those of \citet{Williams2015}.

\onecolumn
\begin{threeparttable}
\caption{All candidate HMXBs in M33 identified in this work. The T11 identifier is in column 1, the corresponding {\it HST} field is listed in column 2, the RA and Dec positions for the X-ray source from T11 are in columns 3-4, column 5 lists the deprojected galactocentric radius in kpc, the source classification from T11 is in column 6, the \citet{Grimm2005} match is in column 8, and any notes on the source are in column 9.}\label{candidate_tab}
\tiny
\begin{tabular}{cccccccc}
\hline \hline
T11 Name & Field & RA & Dec & R$_{gal}$ (kpc) & T11 Class & Grimm Match & Notes \\
\hline
013305.14+303001.4 & 11079-M33-OB127 & 23.27145611 & 30.50041000 & 3.90 & XRB & -- & no {\it HST} counterpart, low SFR \\
013314.68+304012.2 & 9127-M33-PAR-FLD1$^{\bf s}$ & 23.31116694 & 30.67008000 & 3.30 & USNO-0626 & -- & no SFH \\
013315.16+305318.2 & 11079-M33-OB137 & 23.31316694 & 30.88839000 & 5.86 & XRB?,X-4 & J013315.1+305317 & $>$ 3$\sigma$ var \\
013318.34+302840.4 & 5914-NGC598-U137 & 23.32641694 & 30.47789000 & 3.29 & -- & -- & low SFR \\
013320.80+302948.0 & 10190-M33-626sw-28004 & 23.33666694 & 30.49669000 & 2.99 & -- & -- & red counterpart\\
013330.19+304255.6 & 8207-M33-PAR-FIELD4 & 23.37579194 & 30.71547000 & 2.37 & stellar & -- & --\\
013331.32+303402.2 & 6038-M33-AM6-FIELD & 23.38050000 & 30.56728000 & 1.83 & -- & -- & -- \\
013331.86+304011.7 & 10190-M33-301sw-25900 & 23.38275000 & 30.66994000 & 1.77 & -- & -- & low SFR \\
013332.19+303656.8 & 6431-NGC598-FIELD & 23.38412500 & 30.61578000 & 1.54 & stellar & -- & red counterpart \\
013332.23+303955.5 & 8018-M33 & 23.38429194 & 30.66544000 & 1.70 & XRT-2 & -- & -- \\
013332.71+303339.3 & 6038-M33-AM6-FIELD & 23.38629194 & 30.56094000 & 1.83 & -- & -- \\
013334.13+303211.3 & 5998-M33-FLD4 & 23.39223389 & 30.53648000 & 2.02 & XRB,X-7 & J013334.1+303210 & -- \\
013334.54+303556.1 & 6431-NGC598-FIELD & 23.39392611 & 30.59894000 & 1.42 & XRB & -- & -- \\
013336.04+303332.9 & 5998-M33-FLD4 & 23.40018889 & 30.55914000 & 1.69 & -- & J013336.0+303333 & low SFR, $>$ 3$\sigma$ var \\
013336.84+304757.3 & 9873-NGC598-U49 & 23.40350000 & 30.79925000 & 3.05 & -- & -- & low SFR\\
013337.90+303837.2 & 10190-M33-146sw-26505 & 23.40791694 & 30.64369000 & 1.08 & stellar & -- & red counterpart\\
013339.01+302115.0 & 5998-M33-FLD3 & 23.41254194 & 30.35417000 & 4.80 & XRT-3 & -- & --\\
013339.46+302140.8 & 5998-M33-FLD3 & 23.41444694 & 30.36136000 & 4.70 & -- & -- & --\\
013340.09+304323.1 & 9873-M33-267nw-31659 & 23.41705611 & 30.72310000 & 1.68 & -- & -- & -- \\
013340.81+303524.2 & 10190-M33-DISK1 & 23.42004194 & 30.59006000 & 1.17 & -- & -- & --\\
013341.26+303213.4 & 5998-M33-FLD4 & 23.42191694 & 30.53706000 & 1.90 & XRB,XRT-4 & -- & -- \\
013341.47+303815.9 & 10190-M33-146sw-26505 & 23.42279194 & 30.63775000 & 0.78 & -- & -- & -- \\
013341.56+304136.4 & 5384-M33-NEB6 & 23.42320500 & 30.69345000 & 1.17 & -- & -- & -- \\
013342.54+304253.3 & 9873-M33-267nw-31659 & 23.42725000 & 30.71483000 & 1.39 & XRB & -- & no {\it HST} counterpart \\
013344.17+302205.4 & 5998-M33-FLD3 & 23.43404194 & 30.36817000 & 4.70 & -- & -- & --\\
013350.50+303821.4 & 6640-NGC598-SRV2$^{\bf s}$ & 23.46044306 & 30.63930000 & 0.34 & stellar & -- & -- \\
013350.89+303936.6 & 5464-NGC598-1$^{\bf s}$ & 23.46208306 & 30.66017000 & 0.00 & XRB,X-8 & -- & nucleus, no {\it HST} counterpart \\
013351.13+303823.7 & 6640-NGC598-SRV2$^{\bf s}$ & 23.46304194 & 30.63994000 & 0.35 & -- & -- & --\\
013354.28+303347.8 & 6640-NGC598-SRV5 & 23.47617500 & 30.56330000 & 1.77 & L10-069 & -- & SNR? \\
013354.47+303414.5 & 10190-M33-DISK1 & 23.47698000 & 30.57070000 & 1.65 & stellar & -- & --\\
013355.24+303528.6 & 10190-M33-DISK1 & 23.48016694 & 30.59128000 & 1.37 & -- & -- & --\\
013356.77+303729.7 & 6640-NGC598-SRV2$^{\bf s}$ & 23.48654194 & 30.62492000 & 0.93 & QSO/AGN? & -- & $>$ 2$\sigma$ var\\
013356.82+303706.7 & 6640-NGC598-SRV3 & 23.48675000 & 30.61853000 & 1.03 & XRB & -- & red counterpart\\
013358.07+303754.5 & 8207-M33-PAR-FIELD7 & 23.49197889 & 30.63182000 & 0.93 & L10-078 & -- & SNR? \\
013358.23+303438.2 & 10190-M33-419se-160 & 23.49265194 & 30.57730000 & 1.76 & stellar & -- & --\\
013358.50+303332.2 & 10190-M33-419se-160 & 23.49378000 & 30.55896000 & 2.06 & L10-081 & J013358.4+303333 & SNR?\\
013400.75+303944.9 & 5384-M33-NEB1$^{\bf s}$ & 23.50313694 & 30.66248000 & 0.87 & -- & -- & --\\
013401.16+303242.3 & 10190-M33-419se-160 & 23.50484111 & 30.54511000 & 2.44 & stellar & -- & red counterpart, $>$ 2$\sigma$ var\\
013402.37+303136.2 & 10190-M33-419se-160 & 23.50990389 & 30.52674000 & 2.81 & -- & -- & --\\
013402.86+304151.2 & 5914-NGC598-R14$^{\bf s}$ & 23.51191694 & 30.69756000 & 1.00 & stellar & J013402.8+304151 & red counterpart, $>$ 3$\sigma$ var \\
013407.63+303902.4 & 5914-NGC598-R12 & 23.53179194 & 30.65067000 & 1.56 & stellar & -- & --\\
013408.32+303851.7 & 5914-NGC598-R12 & 23.53466694 & 30.64772000 & 1.65 & -- & -- & --\\
013410.03+302856.2 & 10190-M33-784se-5499 & 23.54179194 & 30.48228000 & 3.98 & -- & -- & low SFR\\
013410.51+303946.4 & 5494-M33-OB6-5-FIELD$^{\bf s}$ & 23.54379194 & 30.66289000 & 1.74 & stellar & -- & no SFH\\
013410.69+304224.0 & 11079-M33-OB94$^{\bf s}$ & 23.54457111 & 30.70667000 & 1.63 & L10-096 & -- & SNR? \\
013416.50+305156.5 & 11079-M33-OB39S$^{\bf s}$ & 23.56878306 & 30.86571000 & 3.30 & stellar & J013416.3+305154 & --\\
013417.08+303426.6 & 11079-M33-OB101$^{\bf s}$ & 23.57116694 & 30.57408000 & 3.20 & stellar & -- & $>$ 3$\sigma$ var \\
013417.17+304843.8 & 11079-M33-OB77$^{\bf s}$ & 23.57154194 & 30.81217000 & 2.70 & -- & -- & --\\
013421.15+303930.7 & 5494-M33-OB6-5-FIELD$^{\bf s}$ & 23.58812500 & 30.65853000 & 2.71 & -- & -- & no SFH\\
013423.86+303847.6 & 9873-NGC598-M9 & 23.59941694 & 30.64656000 & 3.04 & stellar & -- & red counterpart, $>$ 2$\sigma$ var\\
013426.14+303726.6 & 9873-NGC598-M9 & 23.60891694 & 30.62406000 & 3.43 & -- & -- & --\\
013432.60+304704.1 & 5237-M33-FIELDN604 & 23.63586806 & 30.78448000 & 3.48 & stellar & -- & --\\
013435.09+304712.0 & 5237-M33-FIELDN604 & 23.64622000 & 30.78668000 & 3.67 & -- & -- & --\\
013438.89+304117.4 & 11079-M33-OB90 & 23.66204194 & 30.68819000 & 4.10 & -- & -- & --\\
013444.71+303732.9 & 8207-M33-PAR-FIELD2 & 23.68631306 & 30.62581000 & 5.06 & USNO-2664 & -- & --\\
\hline
\end{tabular}
\begin{tablenotes}
      \small
      \item [s] Field depth does not reach 80 Myr MSTO. 
    \end{tablenotes}
  \end{threeparttable}
\clearpage
\twocolumn

\onecolumn
\begin{longtable}{cccccccc}
\caption{The list of candidate optical counterparts for each candidate HMXB in this work. The T11 identifier for the HMXB candidate is in column 1, the {\it HST} magnitudes in the F336W, F439W, F555W, F606W, and F814W filters (where available) are in columns 2-6, and the RA and Dec positions for the candidate optical counterpart from {\it HST} are in columns 7-8.} \label{opt_tab} \\
\hline
   \multicolumn{1}{c}{T11 Name} &
   \multicolumn{1}{c}{m$_{F336W}$} &
   \multicolumn{1}{c}{m$_{F439W}$} &
   \multicolumn{1}{c}{m$_{F555W}$} &
   \multicolumn{1}{c}{m$_{F606W}$} &
   \multicolumn{1}{c}{m$_{F814W}$} &
   \multicolumn{1}{c}{{\it HST} RA} & 
   \multicolumn{1}{c}{{\it HST} Dec} \\

\hline
\endfirsthead
   \hline
   \multicolumn{1}{c}{T11 Name} &
   \multicolumn{1}{c}{m$_{F336W}$} &
   \multicolumn{1}{c}{m$_{F439W}$} &
   \multicolumn{1}{c}{m$_{F555W}$} &
   \multicolumn{1}{c}{m$_{F606W}$} &
   \multicolumn{1}{c}{m$_{F814W}$} &
   \multicolumn{1}{c}{{\it HST} RA} & 
   \multicolumn{1}{c}{{\it HST} Dec} \\
\hline
\endhead
  \hline
\endfoot

\endlastfoot
013305.14+303001.4 & -- & -- & -- & -- & -- & -- & -- \\
013314.68+304012.2 & 19.0 & 20.71 & -- & -- & -- & 23.31123066 & 30.67003304 \\
013315.16+305318.2 & 20.8 & 21.84 & -- & -- & -- & 23.31316801 & 30.88824967 \\
013315.16+305318.2 & 22.12 & 23.16 & -- & -- & -- & 23.31324241 & 30.88826482 \\
013315.16+305318.2 & 21.5 & 23.34 & -- & -- & -- & 23.31323366 & 30.88829689 \\
013315.16+305318.2 & 20.01 & 21.84 & -- & -- & -- & 23.31316791 & 30.88827283 \\
013315.16+305318.2 & 20.15 & 20.6 & -- & -- & -- & 23.31323992 & 30.88841399 \\
013315.16+305318.2 & 18.0 & 19.51 & -- & -- & -- & 23.31324649 & 30.88843833 \\
013318.34+302840.4 & -- & -- & 23.73 & -- & 23.47 & 23.32647866 & 30.47776002 \\
013320.80+302948.0 & -- & -- & -- & 21.0 & 20.14 & 23.33678044 & 30.4967935 \\
013330.19+304255.6 & 22.21 & 23.78 & -- & -- & -- & 23.37579808 & 30.71524818 \\
013331.32+303402.2 & 21.93 & 23.17 & -- & -- & -- & 23.38042429 & 30.56740602 \\
013331.86+304011.7 & -- & -- & -- & 23.87 & 23.72 & 23.38275139 & 30.67001082 \\
013332.19+303656.8 & -- & 22.26 & 21.64 & -- & 20.88 & 23.38404693 & 30.61581187 \\
013332.23+303955.5 & -- & -- & 24.2 & -- & 24.19 & 23.38421691 & 30.66537972 \\
013332.23+303955.5 & -- & -- & 24.31 & -- & 24.17 & 23.38433879 & 30.66532402 \\
013332.71+303339.3 & 22.72 & 23.71 & -- & -- & -- & 23.38628803 & 30.56084971 \\
013332.71+303339.3 & 22.4 & 23.11 & -- & -- & -- & 23.3861842 & 30.56078041 \\
013334.13+303211.3 & 17.84 & 19.33 & -- & -- & -- & 23.39227098 & 30.53648949 \\
013334.54+303556.1 & -- & 23.62 & 23.46 & -- & 23.78 & 23.39396567 & 30.59878273 \\
013336.04+303332.9 & 21.83 & 23.27 & -- & -- & -- & 23.40019675 & 30.55918211 \\
013336.84+304757.3 & -- & -- & -- & 23.97 & 24.05 & 23.40356797 & 30.79936029 \\
013337.90+303837.2 & -- & -- & -- & 21.72 & 21.23 & 23.4078521 & 30.64373483 \\
013339.01+302115.0 & 22.28 & 23.84 & -- & -- & -- & 23.41247503 & 30.35454243 \\
013339.01+302115.0 & 22.62 & 23.82 & -- & -- & -- & 23.4129489 & 30.35427171 \\
013339.46+302140.8 & 20.68 & 22.2 & -- & -- & -- & 23.41426396 & 30.36155868 \\
013339.46+302140.8 & 20.11 & 21.22 & -- & -- & -- & 23.41471594 & 30.36120084 \\
013340.09+304323.1 & -- & -- & -- & 23.22 & 23.03 & 23.41715764 & 30.72309722 \\
013340.81+303524.2 & -- & -- & -- & 22.92 & 22.96 & 23.42018113 & 30.59010136 \\
013340.81+303524.2 & -- & -- & -- & 23.74 & 23.75 & 23.41984718 & 30.59000306 \\
013340.81+303524.2 & -- & -- & -- & 23.62 & 23.51 & 23.42012371 & 30.58999796 \\
013341.26+303213.4 & 20.89 & 22.11 & -- & -- & -- & 23.42179328 & 30.53689766 \\
013341.26+303213.4 & 20.55 & 22.07 & -- & -- & -- & 23.42200047 & 30.53714006 \\
013341.47+303815.9 & -- & -- & -- & 21.76 & 21.79 & 23.422621 & 30.6377089 \\
013341.56+304136.4 & 19.26 & 19.40 & -- & -- & -- & 23.422970 & 30.693422 \\
013342.54+304253.3 & -- & -- & -- & -- & -- & -- & -- \\
013344.17+302205.4 & 20.16 & 21.42 & -- & -- & -- & 23.43386426 & 30.36753758 \\
013350.50+303821.4 & -- & -- & 21.0 & -- & 21.14 & 23.4605127 & 30.6394515 \\
013350.89+303936.6 & -- & -- & -- & -- & -- & -- & -- \\
013351.13+303823.7 & -- & -- & 23.9 & -- & 23.68 & 23.46296903 & 30.63977172 \\
013351.13+303823.7 & -- & -- & 23.12 & -- & 23.01 & 23.46308108 & 30.63981074 \\
013354.28+303347.8 & -- & -- & 20.62 & -- & 20.78 & 23.47624293 & 30.56341926 \\
013354.28+303347.8 & -- & -- & 23.3 & -- & 23.22 & 23.47612905 & 30.56351518 \\
013354.47+303414.5 & -- & -- & -- & 24.22 & 24.12 & 23.47717516 & 30.57069537 \\
013354.47+303414.5 & -- & -- & -- & 22.59 & 22.71 & 23.47681386 & 30.57079985 \\
013355.24+303528.6 & -- & -- & -- & 23.91 & 23.89 & 23.48011294 & 30.59113955 \\
013356.77+303729.7 & -- & -- & 21.34 & -- & 21.24 & 23.48658666 & 30.62479502 \\
013356.82+303706.7 & -- & -- & 22.18 & -- & 21.29 & 23.48681041 & 30.6184458 \\
013358.07+303754.5 & 22.02 & 23.27 & -- & -- & -- & 23.49218916 & 30.63184171 \\
013358.07+303754.5 & 22.73 & 23.97 & -- & -- & -- & 23.49230198 & 30.63170482 \\
013358.23+303438.2 & -- & -- & -- & 23.89 & 23.92 & 23.49263863 & 30.57743719 \\
013358.23+303438.2 & -- & -- & -- & 23.7 & 23.71 & 23.49267909 & 30.5772012 \\
013358.50+303332.2 & -- & -- & -- & 24.15 & 23.88 & 23.49362144 & 30.55902244 \\
013358.50+303332.2 & -- & -- & -- & 22.75 & 22.83 & 23.49366451 & 30.55899361 \\
013358.50+303332.2 & -- & -- & -- & 23.25 & 23.29 & 23.49389244 & 30.55884499 \\
013358.50+303332.2 & -- & -- & -- & 23.49 & 23.49 & 23.4938756 & 30.55890179 \\
013358.50+303332.2 & -- & -- & -- & 23.85 & 23.68 & 23.49375081 & 30.55915164 \\
013400.75+303944.9 & 21.55 & 22.76 & -- & -- & -- & 23.50302704 & 30.66263685 \\
013401.16+303242.3 & -- & -- & -- & 22.32 & 21.34 & 23.50485367 & 30.54517402 \\
013402.37+303136.2 & -- & -- & -- & 23.96 & 23.89 & 23.50993725 & 30.52692941 \\
013402.86+304151.2 & -- & -- & 22.44 & -- & 22.05 & 23.51192991 & 30.6977246 \\
013407.63+303902.4 & -- & -- & 23.74 & -- & 23.73 & 23.53177081 & 30.65053608 \\
013407.63+303902.4 & -- & -- & 23.95 & -- & 23.95 & 23.53186385 & 30.65073564 \\
013407.63+303902.4 & -- & -- & 23.79 & -- & 23.52 & 23.53161125 & 30.6505667 \\
013408.32+303851.7 & -- & -- & 23.67 & -- & 23.82 & 23.53472254 & 30.64788787 \\
013408.32+303851.7 & -- & -- & 23.3 & -- & 23.39 & 23.5348752 & 30.64767036 \\
013410.03+302856.2 & -- & -- & -- & 23.83 & 23.7 & 23.54181895 & 30.48218697 \\
013410.51+303946.4$^{\bf x}$ & -- & -- & 22.19 & -- & -- & 23.543699 & 30.662796 \\
013410.69+304224.0 & 21.21 & 22.78 & -- & -- & -- & 23.54455387 & 30.7066734 \\
013416.50+305156.5 & 20.51 & 20.54 & -- & -- & -- & 23.56864113 & 30.86553403 \\
013416.50+305156.5 & 21.25 & 22.39 & -- & -- & -- & 23.56905053 & 30.86585631 \\
013417.08+303426.6 & 20.13 & 21.13 & -- & -- & -- & 23.57127714 & 30.57405008 \\
013417.08+303426.6 & 19.64 & 21.0 & -- & -- & -- & 23.57106813 & 30.57402997 \\
013417.17+304843.8 & 20.23 & 22.03 & -- & -- & -- & 23.57150248 & 30.81218826 \\
013421.15+303930.7$^{\bf x}$ & -- & -- & 22.15 & -- & -- & 23.58817549 & 30.6586403 \\
013423.86+303847.6 & -- & -- & -- & 22.45 & 21.82 & 23.59941327 & 30.6465382 \\
013426.14+303726.6 & -- & -- & -- & 23.74 & 23.77 & 23.60887814 & 30.62418809 \\
013432.60+304704.1$^{\bf +}$ & -- & -- & 19.84 & -- & 19.6 & 23.63573839 & 30.78461361 \\
013435.09+304712.0 & -- & -- & 23.75 & -- & 23.65 & 23.64633193 & 30.78652903 \\
013438.89+304117.4 & 22.11 & 23.67 & -- & -- & -- & 23.66212068 & 30.68814925 \\
013438.89+304117.4 & 22.18 & 23.5 & -- & -- & -- & 23.66208939 & 30.68815607 \\
013444.71+303732.9 & 21.03 & 22.33 & -- & -- & -- & 23.68623071 & 30.62583536 \\
013444.71+303732.9 & 19.6 & 21.13 & -- & -- & -- & 23.68636008 & 30.62576648 \\
\hline
\end{longtable}
\begin{minipage}{\linewidth}
\renewcommand{\footnoterule}{}
{\myfnmark{+} Two saturated sources within the error circle are not recovered in the {\it HST} photometry, so only the faintest candidate counterpart in the error circle is listed here. The LGGS magnitudes of the saturated sources are listed in Section~\ref{sec:indsources}. }\\
{\myfnmark{x} Optical counterpart only recovered in one filter; color cuts made on the basis of LGGS photometry. }
\end{minipage}
\twocolumn

As noted in Section~\ref{classify}, 11 $\pm$ 3 out of the 55 HMXB candidates may be misclassified due to a chance coincidence between the X-ray error circle and a bright, blue star. In column 9 of Table~\ref{candidate_tab} we list some of the source properties for each candidate HMXB, which may be used to cull the population down to the most reliable candidates. For the purposes of the analysis in Section~\ref{sec:agedist} we remove only those sources present in shallow fields (i.e. field depth not sensitive to MSTO at 80 Myr), as these sources will have large errors towards old ages (11 sources), as well as those sources for which there were too few stars in the field to recover a SFH (3 sources). Culling on these sources removes 14 sources, bringing the number of HMXB candidates above our field depth cuts to 41 sources. In Section~\ref{sec:agedist} we present the age distribution for the full sample including sources in shallow fields (52 sources), as well as this culled sample (41 sources).

Using the {\it HST} photometry and the color and magnitude cuts described in Section~\ref{classify} we also ascribed a most likely optical counterpart candidate to each candidate HMXB. For each HMXB candidate we list the likely optical counterpart candidate's position in the aligned {\it HST} catalog and magnitudes in all available {\it HST} filters in Table~\ref{opt_tab}. For some sources there was more than one bright source that passed our color and magnitude cuts within the X-ray error circle, in which case we list the multiple candidate counterparts to the X-ray source. It is impossible to reliably determine any candidate counterpart for the nucleus of M33, so there are no counterparts listed for this source (013350.89+303936.6) in Table~\ref{opt_tab}. If a source was denoted as ``XRB" in T11, but we found no optical counterpart in the {\it HST} images within the X-ray error circle it is still included in Table~\ref{opt_tab} but without a counterpart listed. For these sources it is possible that the {\it HST} filters available (e.g. F336W) were too shallow and blue to pick up optical counterparts that might otherwise be visible in ground-based imaging or with IR imaging for more heavily extincted sources.

\subsection{Individual Sources}\label{sec:indsources}

While most of our candidates are either previously unclassified, or previously classified as associated with bright stars some have interesting history in the literature. Below we present notes on individual sources of interest, including any prior source classifications. 

\begin{figure}
\begin{minipage}[b]{\linewidth}
\centering%
\includegraphics[width=\linewidth,trim=20 0 20 20, clip]{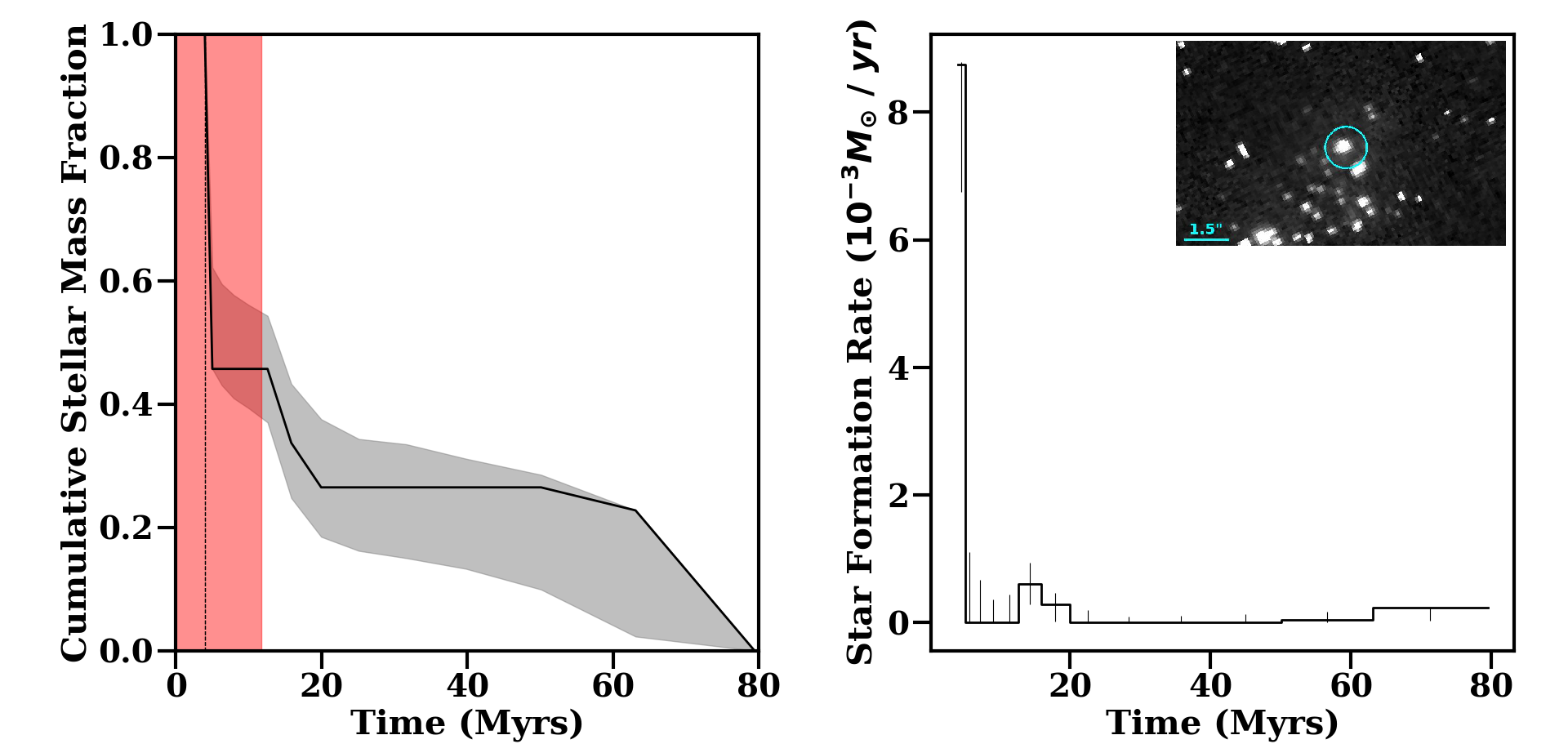}
\end{minipage}
\caption{{\it Left}: The cumulative stellar mass fraction formed as a function of time for the stars in the vicinity of the HMXB 013334.13+303211.3 (M33 X-7). The most likely age  ($<$ $\sim$ 5 Myr) is denoted by the dashed line, with narrowest 68th percentile confidence interval in red. The Monte Carlo derived errors on the cumulative stellar mass fraction are denoted in grey. {\it Right}: The corresponding SFH for M33-X7. {\it Right-inset}: F457W WFPC2 image of M33-X7, with the {\it Chandra} error circle (0.7") shown in cyan.}
\label{m33x7_csf}
\end{figure}

{\it 013334.13+303211.3 : M33 X-7}: This source is the known eclipsing X-ray binary in M33 consisting of a 15.65 M$_{\odot}$ black hole and a $\sim$ 70 M$_{\odot}$ O star secondary in a 3.45 day orbit \citep{Pietsch2004,Pietsch2006,Orosz2007}. The source has been well-observed and modeled, with binary evolution models suggesting it could form within $\sim$ 4 Myr \citep{Valsecchi2010}. We identified a unique optical counterpart within the X-ray error circle ($\sim$ 0.7") for this source with an F336W magnitude of 17.84  and a F439W magnitude of 19.33. The SFH in the vicinity of M33 X-7 is shown in Figure~\ref{m33x7_csf}, with an inset showing the location of the optical counterpart secondary within the X-ray error circle. The SFH recovered for this source yields a most likely age of $<$ 5 Myr, with the narrowest 68th percentile confidence interval encompassing an age up to 11.8 Myr, consistent with the binary components, and the evolution models.

\begin{figure*}
\begin{minipage}[b]{.5\linewidth}
\centering%
\includegraphics[width=\linewidth,trim=0 0 0 0, clip]{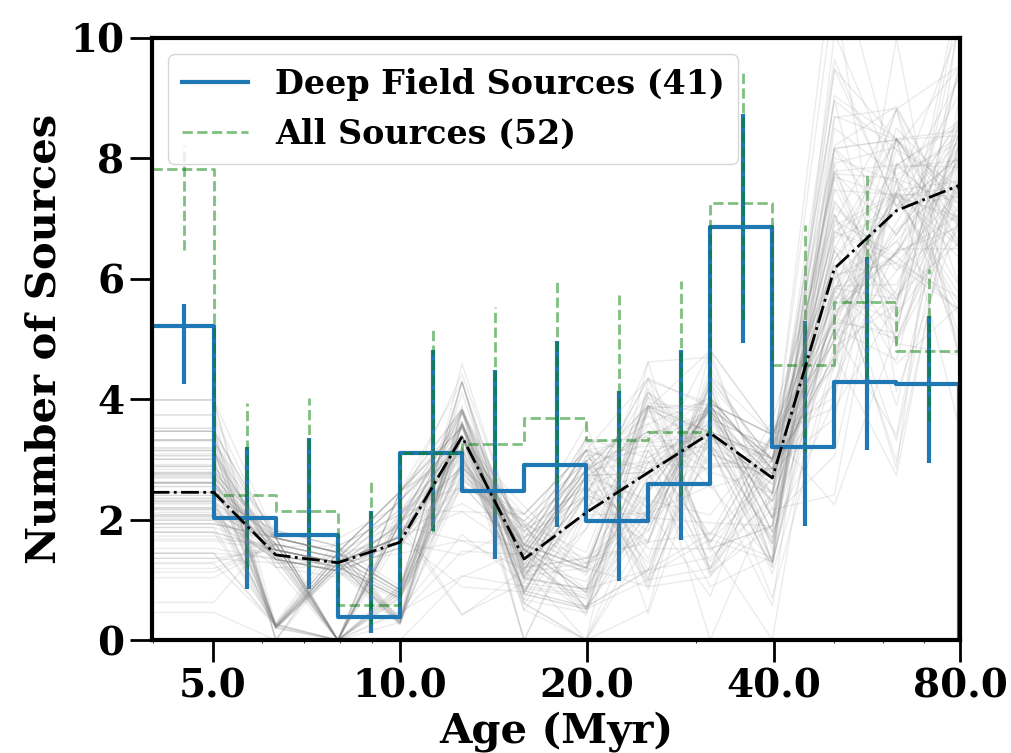}
\end{minipage}%
\hfill%
\begin{minipage}[b]{.5\linewidth}
\centering%
\includegraphics[width=\linewidth,trim=0 0 0 0, clip]{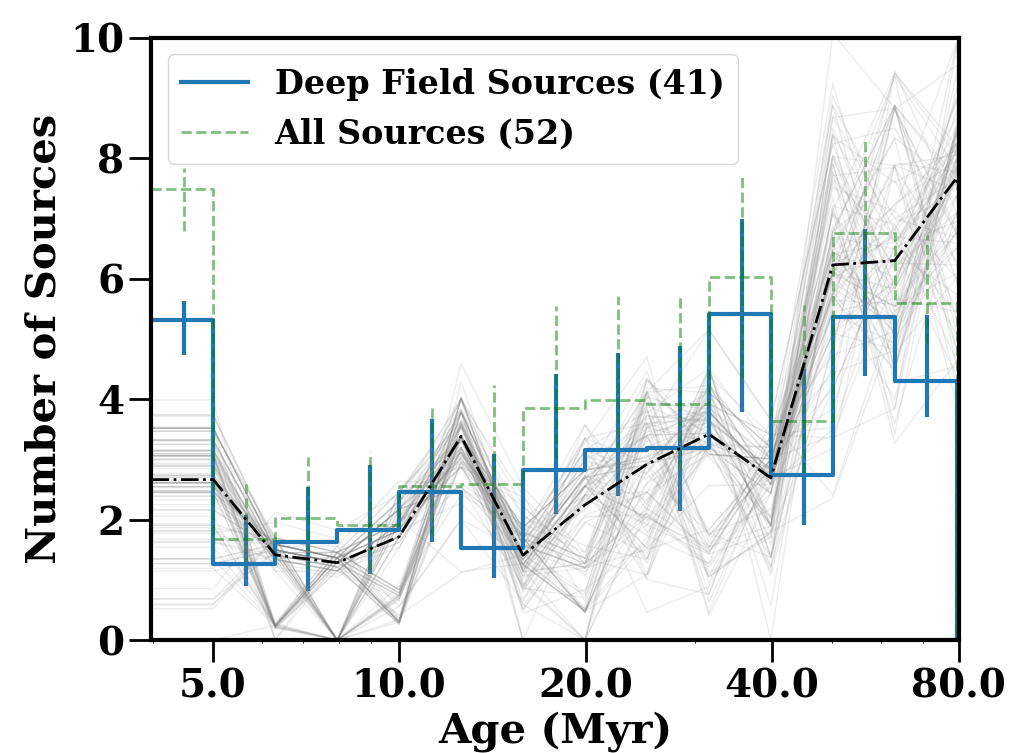}
\end{minipage}
\hfill%
\caption{{\it Left}: The age distribution calculated using 50 pc regions for all 41 HMXBs in M33 that are in fields sensitive to the MSTO at 80 Myr plotted as a solid blue line. The solid blue histogram represents the number of systems expected in each age bin given the SFHs for all sources. The histogram is derived from summing the normalized stellar mass probability distribution functions for each source in each time bin. The distribution is normalized for each source as the sum of the stellar-mass weighted star formation in each time bin where the total mass for each candidate summed to 80 Myr. The errors on the number of candidates in each bin are the Monte Carlo derived errors on the star formation weighted by the stellar mass in each age bin. The light grey lines are the control sample: the equivalent age distribution for 100 random draws of 41 non-HMXB X-ray sources in M33, with the median of these 100 draws as a dash-dot line in black. The HMXB age distribution shows two peaks in HMXB formation efficiency that stand out from the control sample: a peak at very young ages ($<$ 5 Myr), and one at older ages ($\sim$ 40 Myr). The age distribution for all 52 sources in the sample for which we were able to measure SFHs is overplotted as a light green dashed line for reference. {\it Right}: The age distribution calculated using 100 pc regions for all 41 sources in fields sensitive to the MSTO at 80 Myr as a solid blue histogram, with the age distribution for all 52 sources in the sample for which we were able to measure SFHs plotted again as a dashed green line. The light grey lines are the control sample as before.}
\label{agedist}
\end{figure*}

{\it 013354.28+303347.8}: This source was classified as an SNR in \citet{Long2010} (L10-069). Although its X-ray colors are consistent with a soft source (e.g. SNR), it has two bright, blue potential optical counterparts within the X-ray error circle (1.3"). Even with a much smaller error circle (assuming the source is not extended), both counterparts are well within the X-ray position uncertainty. The brightest source within the X-ray error circle has a F555W magnitude of 20.62 and and F814W magnitude of 20.78. Although rare, HMXBs found residing in their natal SNR are not unknown \citep[e.g.][]{Heinz2013}. The narrowest 68th percentile confidence interval suggests a progenitor mass between 8--13.4 M$_{\odot}$, consistent the mass and age derived in \citet{Jennings2014}, and a core-collapse progenitor. We note that the measured SFH also has a peak in the SFR in one of the youngest time bins, indicating that the source could have an age as young as $\sim$ 6~Myr, suggestive of a progenitor $>$ 50 M$_{\odot}$. Further spectral analysis is necessary to determine if an HMXB truly does reside within this SNR, and to discern the nature of the optical counterpart.

{\it 013358.07+303754.5}: This source was classified as an SNR in \citet{Long2010} (L10-078). The X-ray colors for this source are consistent with it being quite soft, but it has two bright, blue stars within the X-ray error circle (1.1"). We note that a significantly smaller error circle would not include these same bright optical candidates. If this source is indeed just an SNR, with no embedded HMXB, the age from its SFH suggests a progenitor mass $>$ 9.7 M$_{\odot}$, which is consistent with a core-collapse supernova progenitor. 

{\it 013358.50+303332.2}: This source was classified as an SNR in \citet{Long2010} (L10-081), but an HMXB candidate in \citet{Grimm2005}. This source is also included in the SNR catalog of \citet{Garofali2017}, where it was found to have X-ray HR values that did not follow the expected trend for an SNR of its temperature, hydrogen column density, and abundance ratio values based on simulated HRs assuming an appropriate SNR spectral model. There are five candidate optical counterparts from {\it HST} within the X-ray error circle (0.7 "). If this source is an SNR the age from its SFH suggests a progenitor mass $>$ 11.7 M$_{\odot}$, consistent with classification as a core-collapse supernova by \citet{Jennings2014,LeeLee2014}.

{\it 013410.69+304224.0}: This source was classified as an SNR in \citet{Long2010} (L10-096). Its X-ray colors from T11 are consistent with it being a soft X-ray source, but there is one bright, blue star within the X-ray error circle (0.7"). If this source is an SNR the age from its SFH suggests a progenitor mass of $>$ 12.4 M$_{\odot}$, which suggests a core-collapse progenitor. This source is in one of the shallow fields, so the young age and high progenitor mass may be the result of limited field depth. 

{\it 013432.60+304704.1}: This source was not previously classified as an HMXB candidate, but is located in the giant H II region NGC 604 in M33. The only candidate optical counterpart recovered from the {\it HST} photometry has an F555W magnitude of 19.84 and an F814W magnitude of 19.6 (as listed in Table~\ref{opt_tab}), however there are two other bright stars visible within the error circle that are saturated, and thus not recovered in the {\it HST} photometry. The sources within the X-ray error circle are blended in the LGGS imaging, but photometry is catalogued for two. These are J013432.63+304704.5 (V magnitude = 17.39, B-V=0.022), and J013432.64+304704.0 (V magnitude=17.19, B-V=0.006), neither of which have confirmed spectral types in the LGGS \citep{Massey2006}. The SFH for this source suggests an extremely young age of $<$ 5 Myr, consistent with the most massive possible compact object progenitor in our models ($>$ 50 M$_{\odot}$), if it is indeed an HMXB.

\subsection{HMXB Age Distribution}\label{sec:agedist}

Resolved SFHs in the vicinity of individual sources allow for constraining HMXB candidate ages, and thus the relevant timescales for feedback, particularly from massive stars and their end-products. Using the CMD fitting technique described in Section~\ref{cmdfit} we measured ages, and thus formation timescales for individual sources, and also built up a distribution of ages for the entire HMXB source population in M33. 

We constructed the age distribution for all 41 candidate HMXBs in fields with depth sensitive to the MSTO at 80 Myr. The full age distribution is the sum of the stellar-mass weighted star formation in each time bin, normalized so that the total mass for each candidate summed to 80 Myr is one. Summing these normalized probability functions
yields the number of expected candidate HMXBs in each time bin. The errors on the number of candidates in each bin are the Monte Carlo derived errors on the star formation weighted by the stellar mass in each age bin.  Stacking the normalized probability distribution functions for all sources in this way makes use of the entire SFH for each source, and thus does not rely on calculating specific ages for individual sources as is described in Section~\ref{technique}. Instead, the stacked distribution gives a measure of the likely HMXB formation timescales based on the SFHs from all sources included in the population analysis. 

The HMXB age distribution for the subsample of 41 sources that pass our field depth cuts is displayed in the left-hand panel Figure~\ref{agedist} as the solid blue histogram, with the distribution for all 52 sources for which we were able to measure SFHs plotted as a dashed light green line. These results are based on using an extraction region of 50~pc for recovering the SFHs, as discussed in Section~\ref{sizeselect}. Doubling the extraction radius resulted in the same age peaks discussed below, but with an increase in the percentage contribution in the oldest age bins, which we attribute to dilution of the signal. For comparison, we show the age distribution using these 100~pc regions for the subsample of 41 deep field sources, and all 52 sources in the right-hand panel of Figure~\ref{agedist}. For reference, we also overplot as light grey lines 100 realizations of the age distribution for a ``control" sample recovered from SFHs in the vicinities of the same number of non-HMXB sources measured in fields where we did not find any candidate HMXBs (grey regions, Figure~\ref{m33coverage}); the median of these 100 draws is plotted as a dash-dot line in black.

The HMXB age distribution contains two prominent peaks that stand out from the control sample: notably a young population at $<$ 5 Myr and an older population at $\sim$ 40 Myr. Both populations can be reproduced in binary population synthesis models, but the young population at $<$ 5 Myr is unusual, as the only known analogues to such a population are a handful of young SG-XRBs in the LMC \citep{Antoniou2016}, the supernova impostor SN 2010da \citep{Binder2016}, and a number of nearby ULXs \citep{Poutanen2013,Berghea2013}. We discuss possible explanations for the populations at both ages based on host galaxy properties and stellar and binary evolution in Section~\ref{LGenviro} and Section~\ref{evo}. 

The full probability distributions for the age of each individual source are listed in Table~\ref{age_tab} (full version available electronically). For a given source each line displays percentage probability that the source was formed in that particular time bin, as well as the corresponding errors in both directions. While we do not ascribe a particular age to each source the most likely age and associated errors can be recovered from the full probability distributions presented in Table~\ref{age_tab} using the method described in Section~\ref{technique}.

\begin{table*}
\caption{The full SFH probability distribution for all HMXB candidates in M33. Column 1 lists the T11 identifier, column 2 is the left edge of the age bin in Myr, column 3 is the right edge of the age bin in Myr, column 4 is the percentage chance that the source was formed in the corresponding age bin (stellar mass fraction in a given bin), and columns 5-6 are the positive and negative errors on that percentage. The full table is available in electronically.}\label{age_tab}
\begin{tabular}{cccccc}
\hline \hline
T11 Name & Age$_\textrm{\scriptsize left}$ (Myr) & Age$_\textrm{\scriptsize right}$ (Myr) & Probability & Probability$^{+}$ & Probability$^{-}$\\
\hline
013334.13+303211.3 &  0.0 & 5.0 & 0.543 & 0.000 & 0.169 \\
013334.13+303211.3 & 5.0 & 6.3 & 0.000 & 0.187 & 0.000 \\
013334.13+303211.3 & 6.3 & 7.9 & 0.000 & 0.190 & 0.000 \\
013334.13+303211.3 & 7.9 &10.0 & 0.000 & 0.185 & 0.000 \\
013334.13+303211.3 & 10.0 & 12.6 & 0.000 & 0.194 & 0.000 \\
013334.13+303211.3 & 12.6 &15.8 & 0.120 & 0.169 & 0.120 \\
013334.13+303211.3 & 15.8 & 20.0 & 0.072 & 0.171 & 0.072 \\
013334.13+303211.3 & 20.0 & 25.1 & 0.000 & 0.200 & 0.000 \\
013334.13+303211.3 & 25.1 & 31.6 & 0.000 & 0.191 & 0.000 \\
013334.13+303211.3 & 31.6 & 39.8 & 0.000 & 0.200 & 0.000 \\
013334.13+303211.3 & 39.8 & 50.1 & 0.000 & 0.211 & 0.000 \\
013334.13+303211.3 & 50.1 & 63.1 & 0.037 & 0.206 & 0.037 \\
013334.13+303211.3 & 63.1 & 79.4 & 0.228 & 0.000 & 0.190 \\
013341.47+303815.9 & 0.0 & 5.0 & 0.000 & 0.100 & 0.000 \\
013341.47+303815.9 & 5.0 & 6.3 & 0.000 & 0.174 & 0.000 \\
013341.47+303815.9 & 6.3 & 7.9 & 0.246 & 0.093 & 0.246 \\
013341.47+303815.9 & 7.9 & 10.0 & 0.000 & 0.253 & 0.000 \\
013341.47+303815.9 & 10.0 & 12.6 & 0.000 & 0.262 & 0.000 \\
013341.47+303815.9 & 12.6 & 15.8 & 0.000 & 0.285 & 0.000 \\
013341.47+303815.9 & 15.8 & 20.0 & 0.000 & 0.321 & 0.000 \\
013341.47+303815.9 & 20.0 & 25.1 & 0.378 & 0.120 & 0.378 \\
013341.47+303815.9 & 25.1 & 31.6 & 0.000 & 0.334 & 0.000 \\
013341.47+303815.9 & 31.6 & 39.8 & 0.000 & 0.309 & 0.000 \\
013341.47+303815.9 & 39.8 & 50.1 & 0.205 & 0.184 & 0.205 \\
013341.47+303815.9 & 50.1 & 63.1 & 0.000 & 0.306 & 0.000 \\
013341.47+303815.9 & 63.1 & 79.4 & 0.171 & 0.039 & 0.171 \\
\hline
\end{tabular}
\end{table*}

\section{Discussion}\label{discuss}

In this section we discuss the population of candidate HMXBs in M33 in terms of the effect of spurious sources on the results, how the age distribution for candidate HMXBs in M33 compares to different Local Group populations of HMXBs, and how the age distribution relates to massive binary evolution. 

\subsection{Effect of Spurious Sources}\label{spurious}

Because HMXBs and background AGN often look very similar in terms of their X-ray spectra and colors it can be hard to distinguish them in the absence of a clearly identified optical counterpart. Even with optical photometry, however, distinguishing the two classes may not be trivial. Reliable source typing may be achieved with the addition of a spectroscopically confirmed counterpart. In this way, false positives, e.g. spectroscopically identified AGN that appear photometrically like HMXB counterparts, may be weeded out from the sample. Because we lack spectroscopic identifications for the entirety of this sample the effect of false positives on the results must be considered. 

As can be seen in Figure~\ref{agedist}, the age distribution derived from the sample of ``control" sources has star formation spread across all time bins, with increasing contribution from contamination at the latest times post-starburst. The low level star formation across all time bins can be thought of as a ``background," while the inflated star formation rate in the oldest age bins comes from higher contribution from contamination that arises due to the effects of stellar mixing when measuring the SFH around sources that are not in exclusively young regions. Thus the contribution of spurious sources to the full age distribution will be to increase the overall ``background" SFR in all time bins, increase the contribution from contamination at old ages, and inflate the error bars on the SFR in each bin accordingly. We address each of these effects on the overall results in turn. 

Increased contribution from contamination at 80~Myr is not a strong concern for HMXB studies, as formation of HMXBs on this timescale is somewhat unphysical given considerations of single star evolution timescales. Therefore, a large signal in this age bin will be a noticeable contaminant. Inflated contribution from contamination in this latest age bin will also have the effect of diluting the signal in adjacent bins, i.e. weakening any discernible signal between $\sim$ 60-80 Myr, if present. This timescale is also unphysical for the formation of HMXBs, so again we would not expect false positives to mask a signal of true physical significance in this time bin.

The presence of false positives in the sample will also increase the SFR (and associated errors) in all other time bins. The effect of this will be primarily to dilute the true age distribution from the population of interest. False positives are unlikely to preferentially boost the contribution to the full age distribution in one of the time bins $<$ 60 Myr unless they all happen to be closely clustered in one region. Therefore, even with a modest fraction of spurious sources included here we do not expect the results to be affected substantially given that we do not have highly clustered candidates.

\subsection{Comparison with Local Group HMXB Populations and Environment}\label{LGenviro}

Resolved SFHs have been used in a number of other nearby galaxies, including the SMC, LMC, NGC 300, and NGC 2403, to look at the age distribution of the HMXB population. In particular,  a preferred age of $\sim$ 40-55 Myr was found for HMXBs in NGC 2403 and NGC 300 by \citet{Williams2013} using {\it HST} data, and 25-60 Myr (peaked strongly at 42 Myr) for HMXBs in the SMC by \citet{Antoniou2010} using the SFH maps of \citet{HZ2004}. More recently, \citet{Antoniou2016} looked at the age distribution for known HMXBs in the LMC using the SFH maps of \citet{HZ2009}. This analysis revealed a peak in the age distribution between 6-25 Myr for the LMC HMXBs. The most likely explanation for differences in the number of HMXBs present at different times is the underlying SFH of the host region, and namely how the SFR varies in regions of the host galaxy on short timescales, as has been demonstrated in other works, both theoretical and observational \citep[e.g.][]{Antoniou2010,Linden2010,Fragos2013,Antoniou2016}. In this section we discuss the preferred formation timescales for HMXBs in M33 as compared to other Local Group galaxies in the context of stellar evolution timescales and the role of metallicity.

The 40~Myr peak in the age distribution as seen in the SMC, NGC 300, NGC 2403, and now M33 is not unexpected assuming the population is strongly dominated by Be-XRBs containing NS companions, as this age coincides with both the peak rate of core-collapse events forming NS assuming standard models of stellar evolution \citep[e.g.][]{Marigo2008}, as well as an observed peak in B star activity \citep{McSwain2005}. The preferred age for these Be-XRBs populations implies that there is an efficient formation pathway for these systems due to binary evolution (see further discussion in Section~\ref{evo}). As these systems are potential precursors to NS-NS binaries \citep{Tauris2017}, this implies that the immediate precursors to these gravitational wave sources may form very efficiently on these timescales across a range of metallicities from significantly sub-solar (e.g. SMC) to solar (M33). 

The $<$ 5~Myr peak that is seen in M33 is observed for a subset of the LMC HMXBs, and in particular most prominently for the SG-XRBs in the LMC. The existence of such a young population, if it is indeed composed primarily of SG-XRBs, may be influenced by host galaxy metallicity. Both M33 and the LMC are higher metallicity environments than the SMC, and in these higher metallicity environments the line-driven winds of massive stars will generally be stronger than in low metallicity environments such as the SMC \citep{Kudritzki1987}. 

The effect of metallicity on the line-driven winds of massive stars has been invoked as a factor in HMXB formation and evolution in terms of its effects on both mass loss and orbital evolution through loss of angular momentum \citep{Dray2006,Linden2010}. In particular, low metallicity environments where the stellar winds are weaker should produce more massive compact objects, as less mass will be lost from the primary via winds, and also possibly systems with different evolutionary pathways, e.g. tighter orbits, due to differences in how much angular momentum is lost from the system. However, in the case of wind-fed accretion onto compact objects as in SG-XRBs the population may be always in a low-luminosity, and thus undetected state in low metallicity environments where there are weak stellar winds, so it is reasonable to expect that this population would not be prevalent in low-metallicity environments \citep[e.g.][]{Linden2010}.

\begin{figure}
\centering%
\includegraphics[width=\linewidth,trim=0 0 0 0, clip]{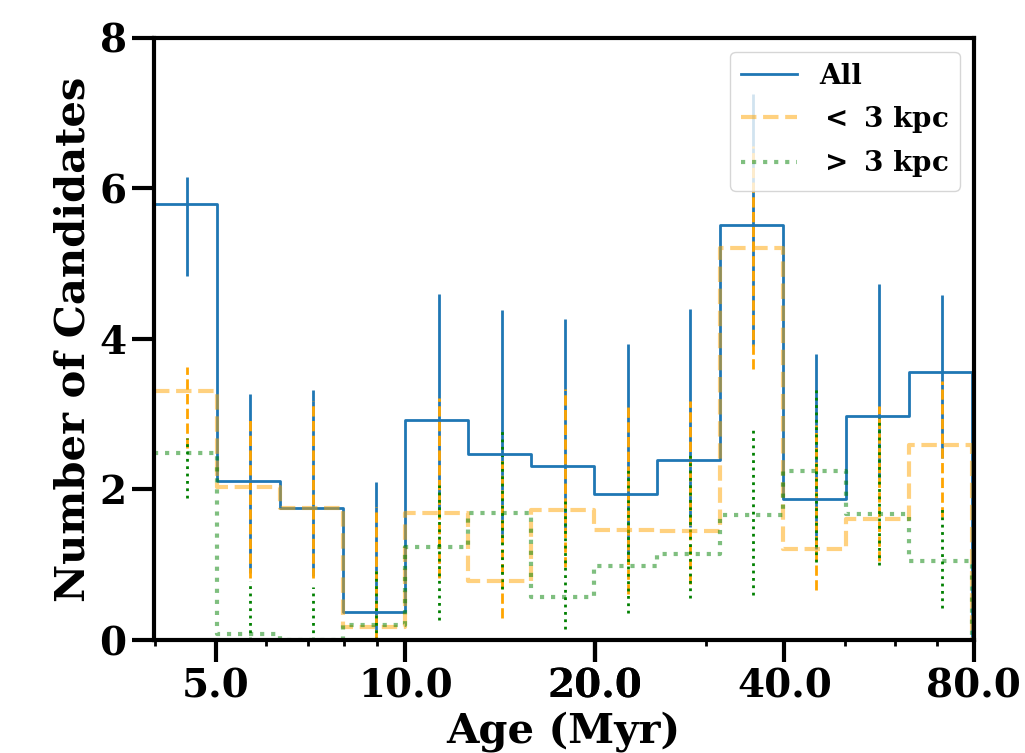}
\caption{The age distribution broken down by galactocentric radius for the 41 candidate HMXBs in M33 that pass our field depth cuts. The full distribution is plotted as a solid blue line, while the distribution for sources within 3 kpc (solar metallicity) of M33's center is plotted as a dashed orange line, and the distribution for sources outside 3 kpc (LMC-like metallicity) is displayed as the dotted green line.}
\label{agedist_breakdown}
\end{figure}

As M33 has a known metallicity gradient which ranges from roughly solar in the interior to LMC-like at larger galactocentric radii with a break at 3 kpc \citep{Magrini2007} we tested for any further metallicity effects by constructing the candidate HMXB age distribution separately for all 41 sources that pass our field depth cuts with galactocentric radii $<$ 3 kpc (dashed orange line), and those with radii $>$ 3 kpc (dotted green line) as shown in Figure~\ref{agedist_breakdown}. Although there appear to be fewer 40 Myr sources at radii $>$ 3 kpc, the small number of sources at these larger galactocentric radii means that we cannot determine any significant differences between the age distributions of the higher metallicity (inner) versus lower metallicity (outer) regions of the galaxy. 

While there are too few sources to make a definite statement about the differences between the candidate HMXBs at solar versus LMC metallicities we can note that there is not a strong preference for young sources originating in one region of the galaxy. In particular, we find that there is no significant difference between the spatial distribution of young HMXB candidates and candidates found at older ages when compared to supergiants in M33 taken from the catalog of \citet{Massey2016}. If we expect that SG-XRBs dominate the younger population in M33 as is the case for the LMC these systems may only require a metallicity threshold (roughly $\frac{1}{3}$ solar) to be observable as bright X-ray sources.

\subsection{Implications for Massive Binary Evolution}\label{evo}

Ages derived from resolved SFHs can help constrain binary evolution scenarios by restricting the formation timescale for HMXBs of various subtypes. There are three main features to the age distribution recovered for HMXB candidates in M33, namely a peak in the number of sources at intermediate ages ($\sim$ 40 Myr), a peak in the number of sources at very young ages ($<$ 5 Myr), and a ``valley" where few, to no sources exist ($\sim$ 6-10 Myr). We discuss each of these features in turn, and their implications for binary evolution. 

\subsubsection{Delayed Onset HMXBs: Be-XRBs}

The peak in HMXB formation around $\sim$ 40 Myr has already been attributed to Be-XRBs, both observationally and based on theoretical understanding of binary evolution \citep{Linden2009,Antoniou2010,Williams2013}. As explained in \citet{Williams2013}, Be-XRBs form on these timescales based on neutron star formation and B star mass loss activity. 

In these systems the initially more massive star (the primary) evolves away from the main sequence and expands to fill its Roche Lobe, in the process transferring mass to its companion (the secondary). The more massive star will end its life first, leaving behind a compact object, most often a neutron star. If the binary survives the death of the primary the result is a secondary that has been spun up by mass transfer that is now orbiting a compact object. Assuming enough material has been transferred such that the secondary is now rapidly rotating it may either liberate material from its poles, which can be funneled down towards the equator to form an equatorial disk, or otherwise directly liberate material around its equatorial regions if its rotation is close enough to the critical velocity \citep{Bjorkman1993}. In either case, an equatorial disk of material is formed around the star, making the star an active Be star \citep{Porter2003}. As mass from this outflowing disk is transferred back to the compact object the system becomes bright in X-rays, and observable as a Be-XRB. 

There are several aspects to the formation of Be-XRBs described above that make the roughly 40 Myr timescale for the appearance of Be-XRBs preferential. Observationally B stars have been shown to have a peak in their activity (the Be phenomenon) between $\sim$ 25-80 Myr, implying that a majority of these stars are spun up by mass transfer in a binary on these timescales \citep{McSwain2005}. This also fits with the preferred timescale for neutron star formation. Assuming a standard Kroupa IMF and a minimum mass threshold for compact object formation, there will be an abundance of neutron stars that form from stars $\sim$ 8 M$_{\odot}$. Assuming standard stellar isochrones this mass corresponds roughly to a 40 Myr timescale for formation of neutron stars in the greatest numbers \citep{Marigo2008}. 

We therefore expect Be-XRBs to be prevalent on this 40 Myr timescale given the above formation scenario, with the caveat that the binary needs to survive the death and supernova explosion of the primary. The binary population synthesis models of \citet{Linden2009} demonstrated that such Be binaries would survive to the HMXB phase if they were preferentially the result of primaries undergoing electron capture supernovae, which would impart smaller kicks to the system. The models of \citet{Linden2009} displayed a prominent peak in the number of HMXBs from $\sim$ 20-60 Myr post-starburst, which they dubbed the electron capture supernovae ``bump," though the strength and age of this bump may be a function of metallicity \citep[e.g.][]{Pod2004}. While our samples do not contain enough sources to definitively compare to their models for electron capture supernovae kick distributions, or pre-explosion core masses, the timescales for formation from simulations are in line with what has been observed here in M33. These HMXB candidates would be those that avoided disruption, and may therefore be made up mainly of systems where the primary underwent an electron capture supernova. 

\subsubsection{Early Onset HMXBs}

The HMXB candidates in M33 with formation timescales $<$ 5 Myr post-starburst are more likely to have compact objects that originate from very massive progenitors based solely on considerations of single star evolution timescales \citep{Marigo2008}. At such young ages, however, the models become somewhat degenerate, so we are not able to ascribe a particular progenitor mass to the stars that form compact objects this early, but rather a lower limit of $\sim$ 50 M$_{\odot}$ for anything formed in the youngest age bins. 

As mentioned in Section~\ref{sec:agedist}, a similarly young population is seen for a small number of interesting sources: the supernova impostor SN 2010da in NGC300 \citep{Binder2016}, a number of nearby ULXs \citep{Poutanen2013, Berghea2013}, as well as a handful of SG-XRBs in the LMC. The formation of SG-XRBs in particular on these timescales is bolstered by binary population synthesis models, like those of \citet{Linden2009}, which predict a large number of massive black holes fed by wind accretion from supergiant companions at early times post-starburst.  

Arguments based on single star evolution would suggest that HMXBs present at very young ages should host black holes as compact objects, as these would be systems originating from the most massive progenitors ($>$ 40 M$_{\odot}$) \citep[e.g.][]{Fryer2012}. There is some evidence for an upper mass cutoff for stars that produce black holes versus neutron stars \citep{Jennings2012,Jennings2014}, however, binary evolution may make it possible for massive progenitors to form neutron stars $\sim$ 5 Myr post-starburst some small fraction of the time if the massive primary undergoes copious amounts of mass transfer or stripping in a binary \citep{BT2008}. Thus, while it is likely that most of the HMXB candidates at these youngest ages host black holes we cannot state so definitively for any source but M33 X-7, for which the mass of the compact object has been measured \citep{Orosz2007}. In the future, {\it NuSTAR} observations of candidate HMXBs in M33 will help delineate between compact object type where dynamical mass measurements are lacking \citep[e.g.][]{Lazzarini2018}. 

While we cannot state with certainty what the nature of the compact object is based on HMXB age alone we can make some inferences bolstered by the combination of system age, candidate optical counterpart color and magnitude, and X-ray luminosity. Based on the {\it HST} magnitudes of the candidate counterparts for the youngest HMXB candidates in M33 the majority of the systems at young ages are likely not systems hosting supergiant donors, though again we cannot say anything definitively about the nature of the compact object. There is likewise no trend with X-ray luminosity and age to indicate a preferred donor--compact object configuration at either young or old HMXB ages. The distribution of X-ray luminosities for sources at young ages ($<$ 5 Myr) is consistent with the luminosity distribution for sources in all other age intervals. 
Only two of the sources in the youngest time bin have candidate optical counterparts with magnitudes brighter than 20th magnitude, sources 013334.13+303211.3 (M33 X-7) and 013432.60+304704.1 (in NGC 604). 

The remainder of the candidate HMXBs in this youngest time bin therefore either represent systems with less massive (possible B star) donors to originally very massive primaries, or otherwise their relatively faint candidate optical counterparts suggest they were in fact formed in later time bins and are therefore older sources, or that the faint photometry of their optical counterpart is the result of past binary interactions (e.g. common envelope phase). 

The possibility that these systems have B star donors to originally very massive primaries is somewhat refuted by current models of binary evolution, which suggest that close binaries with very high mass ratios may not survive the common envelope phase (i.e. would merge), and therefore systems such as Be-BH XRBs should be extremely rare \citep{Belczynski2009}. There is only one such Be-BH binary known in the Milky Way, albeit it is a very low luminosity source \citep{Casares2014,Ribo2017}. 

Instead, it is more likely that these systems may in fact be older, though there is a high probability that they formed during the most recent star formation episode, or that the optical counterpart is fainter than expected due to binary evolution. The expected photometric class of the counterpart given the past history of interactions with a companion is difficult to quantify, as it depends on the unknown initial conditions of the system, as well as uncertain prescriptions for mass loss, mass transfer, common envelope phase, spin-up, and more. We can, however, quantify the likelihood of different formation timescales based on the measured SFHs for each source. In particular, the following four sources do not have candidate supergiant optical counterparts, but have high probabilities of being younger than 5 Myr: 013354.47+303414.5, 013339.01+302115.0, 013330.19+304255.6, 13358.23+303438.2, and 013426.14+303726.6. While the best fits for these sources have most of the stellar mass forming $<$ 5 Myr after the initial burst of star formation there are uncertainties on these best fit values that make it plausible for these sources to have formed in later at later times, and therefore be older. The full SFH probability distributions and their associated errors are listed for each of these sources in Table~\ref{age_tab}. These sources will make interesting candidates for optical follow-up to classify their candidate optical counterparts and {\it NuSTAR} X-ray colors to classify their compact objects. 

\subsubsection{HMXB Valley}

The distribution of sources in M33 shows a distinct lack of sources between the ages of $\sim$ 6-10 Myr. This feature is seen as well, to a certain extent, for HMXBs in the LMC \citep{Antoniou2016}, and is consistent with the binary population synthesis models \citet{Linden2009}, where the HMXBs appearing at the earliest times post-starburst are the systems containing black holes fed by the stellar winds of massive supergiant secondaries, which rapidly disappear once the O stars die out \citep[e.g.][]{vanbever2000}.

However, as shown in Figure~\ref{agedist}, the age distribution for our ``control" sample also shows a steady decrease and minimum between 6-10 Myr, which means that this particular feature is consistent with the SFH elsewhere in M33. Because of this we cannot definitively attribute this feature to binary evolution. It may be that HMXB formation is disfavored on these timescales, but, at least for the case of M33, the SFH also suggests that the star formation rate in general is low on these timescales. 

\section{Conclusions}\label{conclude}

We identified 55 candidate HMXBs in M33 using a combination of deep archival {\it Chandra} and {\it HST} imaging. After precise astrometric alignment between the X-ray and optical data found their candidate optical counterparts. We then used {\it HST} data to construct CMDs in the vicinity of each HMXB candidate. Because HMXBs are young sources still associated with their natal regions we modeled these CMDs to find the best fitting SFHs and probability distribution for the age, or formation timescale of each source. The combination of probability distributions for all sources yields an age distribution for the population of candidate HMXBs, which can be used to infer the preferred formation timescales for HMXBs in M33. 

The age distribution in M33 has three distinct features: a peak in HMXB production at $<$ 5 Myr, a valley where there are few, if any, sources between 6-10 Myr, and another peak in HMXB production at $\sim$ 40 Myr. The 40 Myr peak has been seen in other nearby galaxies, and can be attributed to an efficient formation pathway for Be-XRBs across a range of metallicities. In particular, this peak is expected given increased production of neutron stars from 8 M$_{\odot}$ progenitors on this timescale, possibly via electron capture supernovae, and and a peak in the Be star phenomenon from $\sim$ 25-80 Myr, possibly due to spin up by mass transfer in a binary. The young population seen in M33 is unique compared to what has been observed in other Local Group galaxies, but given binary population synthesis models and stellar evolution timescales it is expected assuming a population of massive progenitors that result in relatively massive black holes accreting from the winds of supergiant companions at very early times (e.g. M33 X-7). The distinct lack of sources between 6-10 Myr may be attributed simply to the particular SFH of M33, which exhibits low a SFR on these timescales.

This sample of candidate HMXBs with measured ages provides not only a population ripe for further observational follow-up, but also a unique set of sources to help place constraints on binary population synthesis models. In particular, this population of candidate HMXBs can help constrain models by requiring them to reproduce the observed HMXB sample not only in terms of numbers of systems, but also when they must be formed relative to the known host galaxy properties of M33. Models that seek to reproduce these candidate HMXBs as an intermediate step in binary evolution can then be used to infer the population of compact object binaries and potential gravitational wave sources that would result by evolving this local population of HMXBs forward in time. 

Future work will expand upon this catalog of HMXB candidates to refine the system characteristics using detailed X-ray spectral fits, and X-ray timing analysis for sources with enough counts. In addition, full spectral energy distribution fitting to determine optical counterpart spectral class will be possible using the upcoming M33 Legacy Survey (PI: Dalcanton). Forthcoming {\it NuSTAR} data for M33 will also yield compact object types from X-ray colors for some candidate HMXBs. Taken together, this will yield the first large sample of HMXBs with known system characteristics in a large spiral galaxy outside the Milky Way. Such a population is of great interest in understanding the potential local analogue precursors to gravitational wave sources. 

\section*{Acknowledgements}

Support for this work was provided by grant GO-14324 from
the Space Telescope Science Institute, which is operated by
the Association of Universities for Research in Astronomy,
Incorporated, under NASA contract NAS5-26555.  We thank the referee for their
very helpful comments in improving the manuscript.








\bsp	
\label{lastpage}
\end{document}